\documentclass[%
reprint,
 superscriptaddress,
showpacs,
 amsmath,amssymb,
 aps,
prb,
]{revtex4-1}

\usepackage{graphicx}% Include figure files
\usepackage{dcolumn}% Align table columns on decimal point
\usepackage{bm}% bold math
\begin{document}
\preprint{Draft9}

%\title{Stiffening of $^4$He and $^3$He Films Adsorbed in Porous Glass}
%\title{Elastic Anomalies in Helium Films at the Quantum Phase Transition}
\title{Elastic Anomaly of Helium Films at a Quantum Phase Transition }

\author{T. Makiuchi}
% \email{tmakiuti@phys.keio.ac.jp}
\affiliation{Department of Physics, Keio University, Yokohama 223-8522, Japan}
 \author{M. Tagai}
 \affiliation{Department of Physics, Keio University, Yokohama 223-8522, Japan}
\author{Y. Nago}
\affiliation{Department of Physics, Keio University, Yokohama 223-8522, Japan}
\author{D. Takahashi}
\affiliation{Center for Liberal Arts and Sciences, Ashikaga University, Ashikaga 326-8558, Japan}
\author{K. Shirahama}
% \email{keiya@phys.keio.ac.jp}
\affiliation{Department of Physics, Keio University, Yokohama 223-8522, Japan}

\date{\today}% It is always \today, today,

\begin{abstract}
 Helium films show various quantum phases that undergo quantum phase transitions by changing coverage $n$.
 We found anomalous elastic phenomena in bosonic $^4$He and fermionic $^3$He films adsorbed on a glass substrate.
 The films stiffen under AC strain at low temperature with an excess dissipation.
 The onset temperature of the stiffening decreases to 0 K as $n$ approaches a critical coverage $n_\mathrm{c}$. 
 The elastic anomaly is explained by thermal activation of helium atoms from the localized to extended states with a distributed energy gap. %anomalies are -> anomaly is 180810巻内
 We determine for the first time the energy band structure of helium films from elasticity.
 The ground states of $^4$He and $^3$He at $n < n_\mathrm{c}$ are identically gapped and compressible, which are possibly %self-organized Mott insulator or Mott glass.%関係代名詞の位置が良くないですが、どうでしょうか。%いいのではないでしょうか？ a self-organized... の冠詞aは消しました．(180808巻内)
 a sort of Mott insulator or Mott glass. %a sort ofに変えました．self-organized MIという提案がこれまでされているわけではないため181015巻内
%了解です181015
\end{abstract}

\maketitle

%\tableofcontents

\section{\label{sec:intro}Introduction}

Quantum phase transition (QPT) has been actively studied in condensed matter physics, because it occurs between emergent quantum phases\cite{SachdevBook}. %quantum を加えて以降消去 by interplay of particle exchange, correlation and disorder. 
In particular, superfluid-- and superconductor--insulator transitions in superconducting films\cite{DobrosavljevicBook} and ultracold atoms in optical potentials\cite{Greiner2002, Bloch2008} are typical examples of QPTs. %a typical example -> typical examplesにしました(180808巻内)
%In SI-QPTs, while superfluid phases are unique and well understood, there are varieties in insulating phases, because the types of insulating phase are determined by competition between quantum fluctuation and other conditions such as interparticle correlation and external potential.%あまりうまくないですが、少し付け足してみました8/7
In superfluid--insulator QPTs, while superfluid phases are unique and well understood, there are various possibilities for insulating phases because they are determined by competition between quantum fluctuations, interparticle correlation, and external potential. %あまりうまくないですが、少し付け足してみました8/7 %このように変えてみましたが，いかがでしょうか？(180808巻内)
%最初の文章のby以降が似ているのでそちらを消しましょう。fluctuationsにしました
In spatially periodic systems such as bosons in optical lattices, the insulating phase is Mott insulator. 
In disordered metals and atoms in disordered potentials, Anderson insulator and Bose glass are the candidates for insulating phases. 
In this paper, we propose that helium films offer a new example of QPT between a superfluid and a novel insulating phase, which has an energy gap and a finite compressibility.
%SIの略をやめました181015巻内
%良いと思います181015

Helium (bosonic $^4$He and fermionic $^3$He) films formed on solid substrates by adsorption undergo various QPTs between competing phases by changing coverage $n$ (areal density) as an external parameter. %show -> undergo 180808巻内　emergentが最初の段落にもあるので消しました
On atomically flat surface of graphite, helium films form clear layer structures from one to several atomic layers\cite{Zimmerli1992}.
Potential corrugation provided by graphite and correlation between helium atoms produce various ordered phases such as Mott insulator, heavy Fermi fluid, nuclear magnetic phases, and coexistence of superfluid and density wave order\cite{Casey2003,Neumann2007,Nyeki2017}. %show -> form, 参考文献Zimmerliを追加，porvided by graphite structureのstructure省略，and nuclear...  -> and削除，coexisting states -> coexistence，文をふたつに分けました 180808巻内
%Correlation and exchange can be controlled by tuning areal density (coverage) of helium atoms, and dimensionality and disorder can also be altered by using porous media as a substrate. 
%Various quantum phases that are not observed in bulk helium are realized.
On disordered substrates, such as glass and Mylar (plastic film), the situation is quite different. %On the other handを省略，誤用と思われます(参考グレン・パケット)．situations are -> the situation is 180808巻内
No clear layer structure is observed, and superfluidity emerges when $n$ exceeds a critical value $n_{\mathrm c}$, which is 6--27 $\mu$mol/m$^2$ (about 0.5--2 atomic layers) depending on substrates\cite{Csathy2003,Crowell1995,*Crowell1997}. %seen -> observed, about追加 180808巻内
%ここでCsathy を引用しますか?(6-28の説明) 引用しました180810巻内
%Crowell引用しました180812
The superfluid transition temperature $T_{\mathrm c}$ increases as $n$ increases from $n_{\mathrm c}$, while films at $n < n_{\mathrm c}$ do not exhibit superfluidity. %coveragesを省略，180808巻内 %while以降を省略181009巻内
The superfluid films undergo a well-known Berezinskii--Kosterlitz--Thouless (BKT) transition on Mylar substrate\cite{Bishop1978,*Bishop1980}, while films in porous media show %\textcolor{magenta}{ [Referee B (3)] (three-dimensional nature of superfluidity$\rightarrow$) %ここはコメントアウトですか?181015 %すみません，やはり入れました181016巻内
both two and three-dimensional characteristics due to the macroscopic connectivity of locally two-dimensional films\cite{Reppy1992,Shirahama1990}. %} %修正181011 critical behavior of superfluidity though the local motion of adsorbed atoms is restricted in two dimensions}\cite{Reppy1992}. %3次元超流動の引用しますか?8/8 -> Reppy1992を足しました 180808巻内 %Referee B's comment (3)の反映181010巻内
%{Shirahama1990}として、私の論文を引用してもらえますか。Size-dependent Kosterlitz-Thouless superfluid transition in thin 4He films adsorbed on porous glasses, K. Shirahama, M. Kubota, S. Ogawa, N. Wada, and T. Watanabe, Phys. Rev. Lett. 64, 1541 (1990)です。181015 %リファーしました181016巻内

The most important feature of $^4$He films on disordered substrates is that there is only one ``quantum critical coverage'' $n_{\mathrm c}$. 
Films at $n < n_{\mathrm c}$ are considered to be in an insulating phase, meaning that a superfluid--insulator QPT occurs at $n_{\mathrm c}$. % is regarded as an -> are considered to be in 180808巻内anを入れました180812
We emphasize that $^4$He on disordered substrates realizes an \textit{ideal} superfluid--insulator QPT. %nearly省略．180808巻内idealというのは意味も判らないので迷いましたがそのままにしてあります。180812
On graphite, $^4$He superfluidity and $^3$He magnetism are strongly influenced by corrugation from substrate. %structure of filmsが不要と感じたので省略しました（少なくともカンマの位置が変？）．periodic corrugation of substrate potential -> corrugation from substrate．180808巻内
It is rather surprising that superfluid transition of $^4$He on Mylar shows a perfect agreement with the BKT theory\cite{Bishop1978,*Bishop1980}, while $^4$He on graphite does not\cite{Crowell1996}.%Bishopの引用入れました180812

%\textcolor{red}{[Referee A] %コメントアウト181013巻内
The existence of $n_\mathrm{c}$ was initially explained by the so-called inert layer model\cite{Chan1974,Washburn1975}.
% This model assumes that  was employed to explain the experimental facts that $^4$He films exhibit superfluidity only when $n$ exceeds $n_\mathrm{c}$, and the superfluid density and transition temperature $T_\mathrm{c}$ are roughly proportional to $n - n_\mathrm{c}$, while at $n < n_\mathrm{c}$ the films is localized and forms a two-dimensional solid. 
%An intuitive and classical picture for the localization and the onset of superfluidity is the one called the inert layer model.
 %In this model, a $^4$He film at a coverage less than $n_\mathrm{c}$ is solid layer(s) completely bound to the substrate. 
In this model, an ``inert'' solid layer adjacent to substrate and a superfluid layer atop the inert layer form two independent subsystems. 
Although this model is consistent with the fact that $n_\mathrm{c}$ depends on helium--substrate potential depth, %substrate material, i.e. helium--substrate potential, 
it does not explain the deviation of $n$--$T_\mathrm{c}$ relation from linearity, and the nonadditivity of heat capacity\cite{Csathy2003}. %エンダッシュの修正，冠詞の追加
% For a film with a coverage $n>n_\mathrm{c}$, the inert layer of a coverage $n_\mathrm{c}$ remains and liquid or superfluid layer of a fraction $(n-n_\mathrm{c})$ floats on it, distant from the strong attraction from the substrate.
% It has been realized, however, that the inert layer model is inadequate to explain experimental results\cite{Reppy1992,Crowell1997,Csathy2003}.
% }
% 
 %At $n < n_{\mathrm c}$, helium adatoms are localized on substrate. %ここのlocalizedを斜体にした理由は何でしょうか？波動関数ではなく，粒子が局在するという古典的な描像の局在，という意味でしょうか？斜体にしなくても良いと思いました．180808巻内強調ですが削除しました %前後の文のつながり上，削除しました181009巻内
%\textcolor{red}{The origin of localization of $^4$He film at $n < n_{\mathrm c}$ is still under survey in the context of the boson localization.} 
%\textcolor{red}{Concentrating on the localization problem of $^4$He film at $n < n_{\mathrm c}$, its major origin is still under survey in the context of the boson localization.}
%The localization is attributed to strong attraction between helium and substrate. %localized state -> localization 180808巻内 %was first -> is 181009巻内
%This conjecture was supported by the fact that $n_\mathrm{c}$ scales with the depths of potential of different substrates\cite{Csathy2003}.
It is therefore desirable to study nature of the localized state and the QPT beyond the inert layer model.%} 

Fisher \textit{et al.} proposed that many-body effects of correlation and disorder make $^4$He film at $n<n_\mathrm{c}$ localized to be a Bose glass, which is characterized by no gap and finite compressibility\cite{Fisher1989}. %in their seminal paper省略．in the localized state省略 180808巻内
% Heat capacity of $^4$He films on porous Vycor glass found no evidence for the Bose glass\cite{Crowell1995,*Crowell1997}.
But no evidence for the Bose glass of $^4$He film was found experimentally \cite{Crowell1995,*Crowell1997}. %上の文を変更180808 wasにしました
We have found anomalous behavior in elasticity of helium films, an important property that is related to a compressibility of ground state. 
The ground state at $n < n_\mathrm{c}$ is found to be a gapped many-body state such as Mott insulator or Mott glass\cite{Giamarchi2001}, which has intermediate properties between Mott insulator and Bose glass. %ここの2文がないと、2つ下の段落で弾性異常がいきなり出てきて、とても唐突に感じます。弾性測定をする理由も述べる必要があります。従ってこの2文を戻した方がよいです。 これがないと、なぜ弾性など測定するのか読者は理解できないと思います %承知しました180810巻内

Contrary to $^4$He, studies of $^3$He films on disordered substrates were few. %veryを省略180808巻内
Since $^3$He films show no superfluidity %\textcolor{blue}{
at currently available low temperatures %}
 and the heat capacity is dominated by a contribution from nuclear spins\cite{Golov1996-2}, critical coverage $n_{\mathrm c}$ was not identified for $^3$He. %atomically thin省略，参考文献Golovを追加．clearlyを省略 180808巻内
%下を変えてみたので、下と一緒の段落の方がわかりやすいです。
Also in $^3$He films, we have observed %elastic anomaly that is identical to $^4$He. % 変えてみました
 the elastic anomaly identical to that of $^4$He films. %前の文では，elastic anomaly = 4Heということになってしまうと思ったので，変更しました．180810巻内
The critical coverage $n_{\mathrm c}$ is identified for the first time for $^3$He.
%The band structure is characterized by energy gap closing as $n$ approaches $n_\mathrm{c}$. 
%showing that the QPT is universal, independent of quantum statistics.省略してみました。
%The critical coverage $n_{\mathrm c}$ of $^4$He film nearly equals to that of the superfluid onset from a previous work \cite{Yamamoto2004}, and that of $^3$He is identified for the first time, showing that the QPT is universal independent of quantum statistics.この文はイントロには必要ないと思います。また長いので3Heで初めてという主張が弱くなっています。載せるならand で2文に分けて下さい。
%前の２つの段落から，我々の貢献を分けて一段落にしました．180808
%またつないでみましたが、検討して下さい。8/10
%こうでないとダメであることがわかりました．どうもありがとうございます．一部分だけ修正しました．180810巻内

\section{\label{sec:expt}Experimental method}

\subsection{Torsional oscillator and porous glass}
\begin{figure}
 \centering
 \includegraphics[width=60mm]{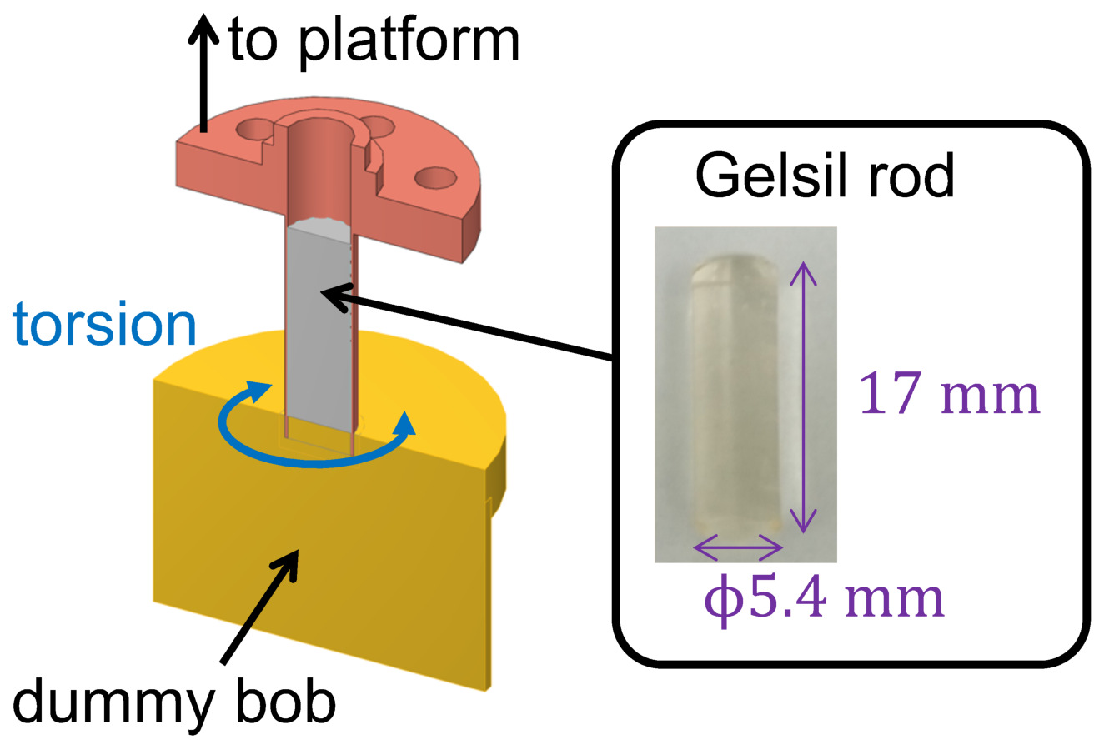}
 \caption{A cross-sectional view of the torsional oscillator. The uppermost part has screw holes for mounting on a platform. The photograph shows the rod sample of porous Gelsil glass we have employed in this work.}
 \label{fig:TO}
\end{figure}

We have measured elasticity of helium films using a torsional oscillator (TO) shown in Fig. \ref{fig:TO}.
%\textcolor{blue}{
Contrary to TO conventionally used in studies of superfluid helium, in which space for helium is located in the bob to measure mass decoupling, %} % 追加しようと思いましたが…181015巻内
%付け加えました181015 %色だけつけました181016巻内 %space.. are->is181016巻内
our present TO consists of a beryllium copper (BeCu) torsion rod containing a porous Gelsil glass sample and a brass dummy bob, which also acts as an electrode for torsional oscillation.  %付加181013
%\textcolor{blue}{
Gelsil is a nanoporous silica glass manufactured by sol-gel method, and has nanopores that are randomly connected. 
Its structure is similar to that of porous Vycor glass, which was typically used in many superfluid helium studies. %} %which acts as a substrate 181010巻内
%GelsilとVycorは製法が異なるので、完全に同じとは言えないと思います181013 削除しました181014巻内
%付け加えてみました181015 %ありがとうございます181016巻内
%\textcolor{magenta}{[Referee B (3)]
Adsorbed helium atoms form a film on the pore wall, and the atoms in fluid state can move along the wall. %} %Referee B's comment (3)の反映181010巻内
The Gelsil sample we used is cylindrical shape, 17 mm in length, and 5.4 mm in diameter.
% The nominal pore diameter is 2.5 nm.下記に移動181013
  %\textcolor{blue}{
  Before the construction %Typo181013
 of the TO, the Gelsil sample was baked at 150 $^\circ$C in vacuum for 3 hours to eliminate adsorbed molecules, especially water. %gases->impurites，さらに二文に分けました． 181015巻内
%impuritiesをmoleculesに替えました181015
 The sample mass after the baking was $0.371$ g. %sample付加181015
 Then we took a nitrogen adsorption--desorption isotherm at 77 K for surface characterization. %} %desorption付加181015 %ありがとうございます．見落としていました181016巻内
 A surface area analyzed with Brunauer--Emmett--Teller method\cite{Brunauer1938} is $166\ \mathrm{m^2}$ (447 $\mathrm{m^2/g}$).
 A pore diameter distribution, analyzed with Barrett--Joyner--Halenda\cite{Barrett1951} method, has a peak at 3.9 nm. %, which is larger than the nominal value. 下にその記述があるので，削除181014巻内
 %\textcolor{blue}{
 This peak pore size is larger than the nominal pore diameter 2.5 nm, which was determined by manufacturer. %これを加えました181013 %ブルーにしました181015巻内
 The Gelsil was again baked for 6 hours, and glued into the BeCu tube (6.0 and 5.5 mm in outer and inner diameter, respectively) with Stycast 1266 epoxy in a $^4$He atmosphere. %} %青字の範囲拡大（in a 4He atmosphereも足したので）181014巻内epoxy修正181015
The TO was mounted on a torsional vibration isolator consisting of a massive copper platform with large rotational moment of inertia (70 mm$\phi$, 30 mm thick) and a copper torsion rod (5 mm$\phi$, 30 mm long). %付加181015
Two brass electrodes, which are for driving and detecting the torsional oscillation, are located on the platform so as to form two parallel plate capacitors with the flat faces of the dummy bob. %付加181013

 The whole TO setup was attached to a cold plate under a mixing chamber of a Joule--Thomson cooled dilution refrigerator (Cryoconcept Inc.). %修正181013 %JT->Joule--Thomson181015巻内
 Sample temperature was measured using a RuO$_2$ thermometer (below 43 mK) and a calibrated germanium thermometer (43 mK--5 K) on the platform.
 The temperature was controlled with %\textcolor{blue}{
 a Manganin twisted wire heater %}
 and the RuO$_2$ thermometer.

\subsection{Finite element method}
For the present TO, the resonant frequency of the torsion mode is simply given by $f=(1/2\pi)\sqrt{k/I}$, where $k$ is a torsion constant (stiffness) of the rod and $I$ is a moment of inertia of the dummy bob.
As $I$ is constant, an increase in $f$ by changing the coverage indicates stiffening of adsorbed film.
Rigorously, however, adsorption of helium on the porous Gelsil glass may slightly contribute to the moment of inertia of the bob, therefore the effects of both stiffening and %weighting 修正181013
mass loading should be investigated.

\begin{figure}[tb]
 \centering
 \includegraphics[width=80mm]{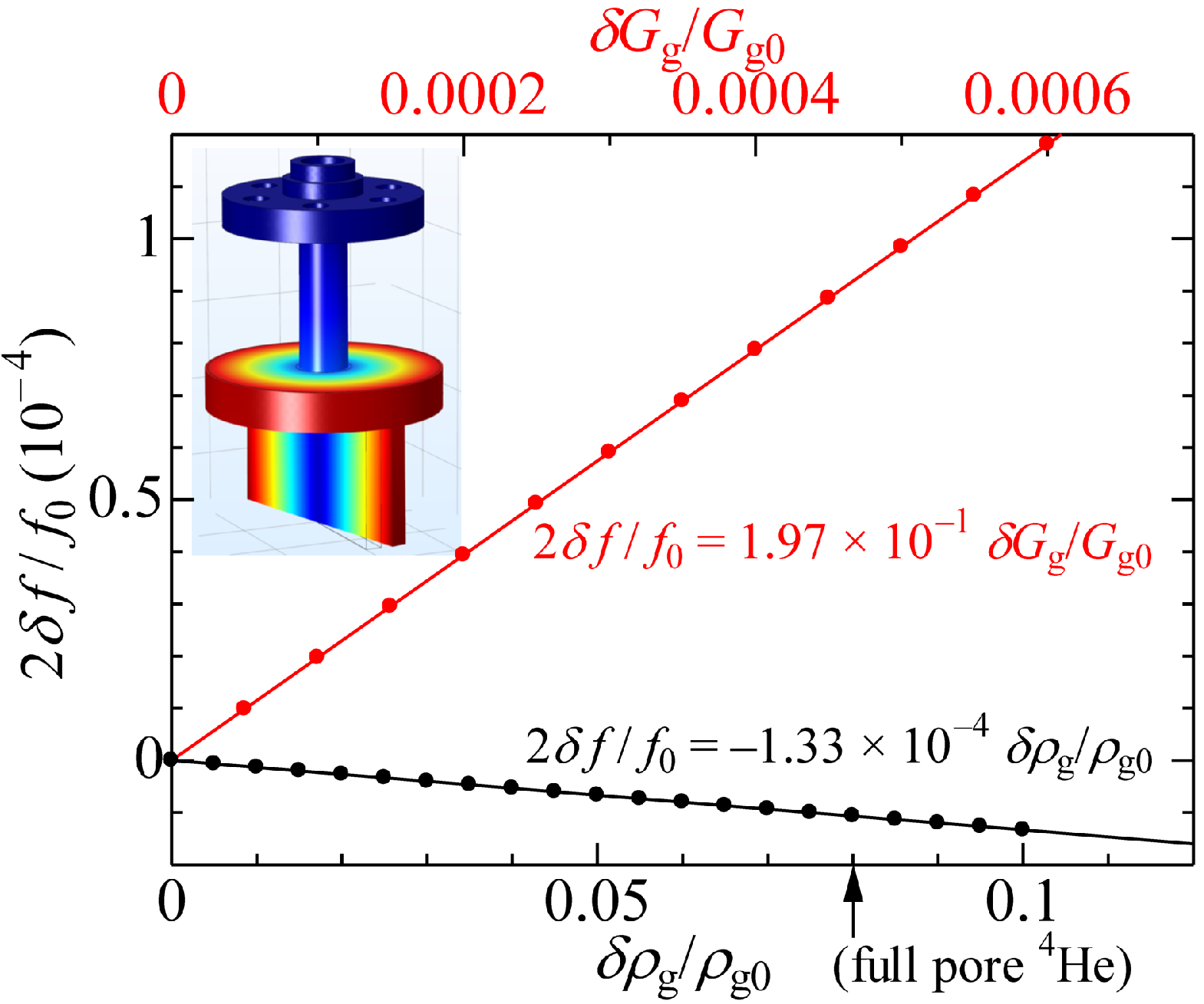}
 \caption{Calculated frequency shift due to changes in the density $\rho_\mathrm{g}$ and in the shear modulus $G_\mathrm{g}$ of the Gelsil sample by FEM simulation.
 A possible maximal value of $\delta \rho_\mathrm{g}/\rho_\mathrm{g0}$ in the case of full-pore $^4$He is indicated with an arrow.
 Inset is a false color picture indicating the movement of parts of our TO. %Insetの説明は?
}
 \label{fig:FEM}
\end{figure}

We performed simulations with finite element method (FEM) to compute how much the resonant frequency of the current TO changes by effective stiffening and mass loading 
%weighting  修正181013
of the Gelsil sample after helium adsorption. %修正181013
We treat the Gelsil rod as a continuous material with a Young's modulus $E=17.1$ GPa and a Poisson's ratio $\nu=0.155$, measured by an ultrasound measurement\cite{Negishi}. 
%(unpublished).は引用に。Y. Negishi and K. Shirahama, unpublished. などでよい。181013
%\cite{[{T. Kogure, H. Yoshimura, R. Higashino, D. Takahashi and K. Shirahama, unpublished}]Kogure}.
 % A 3D CAD model of the TO is drawn with SOLIDWORKS$^{\textregistered}$. The resonant frequencies for several parameters were calculated with COMSOL Multiphysics$^{\textregistered}$ via LiveLink$^{\texttrademark}$ for SOLIDWORKS$^{\textregistered}$.
 The calculated resonant frequency by FEM was $f_0=$ 962 Hz for the present TO, which is about 10 percent larger than the measured value $f_0=$ 860 Hz at low temperatures. %修正181013
The origin of this difference in $f_0$ is not known. %has not been known. 
%If this discrepancy comes from a poor estimate of the torsion constant, we overestimate it by a factor of 1.25. 冗長なため削除(巻内)
%It is too large to explain only by a poor measure of diameter of the torsion rod. 冗長なため削除(巻内)
One possible reason is that the inhomogeneity of silica structure in the porous glass sample, which is not taken into account in the FEM but might affect the resonant frequency in reality. 
%We do not, however, go into detail of the discrepancy because in any case the reduced resonant frequency shift $2\delta f/f_0$ is a good value to consider the influence from Gelsil's physical properties.%この文章の意味がわかりませんが、もう少しうまく説明して下さい。→下のように変えました．（巻内）
The reduced resonant frequency shift $2\delta f/f_0$, however, is a good quantity to compare the measured value to the FEM result.

If the adsorbed helium film stiffens, the apparent shear modulus of Gelsil substrate $G_{\mathrm g}$ will increase, i.e. $G_\mathrm{g}\rightarrow G_\mathrm{g0}+\delta G_\mathrm{g}$. 
Helium adsorption also increases the apparent density of Gelsil, i.e. $\rho_\mathrm{g} \rightarrow \rho_\mathrm{g0} + \delta\rho_\mathrm{g}$. 
%We calculated $f$ when the shear modulus $G_\mathrm{g}$ and the density $\rho_\mathrm{g}$ changes by small factors.
The frequency changes for small $\delta G_\mathrm{g}/G_\mathrm{g0} = \delta E/E$ and $\delta \rho_\mathrm{g}/\rho_\mathrm{g0}$ are well fitted by linear functions as shown in Fig. \ref{fig:FEM}. 
The results are
\begin{equation}
 \frac{2\delta f}{f_0} =  1.97\times 10^{-1} \frac{\delta G_\mathrm{g}}{G_\mathrm{g0}},
 \label{eq:FEMg}
\end{equation}
and
\begin{equation}
 \frac{2\delta f}{f_0} = -1.33\times 10^{-4} \frac{\delta \rho_\mathrm{g}}{\rho_\mathrm{g0}}.
 \label{eq:FEMr}
\end{equation}
%\textcolor{blue}{
The effect of elasticity is larger than that of mass loading 
%weighting 修正181013
by a factor of $10^3$ in the present TO. %}

The effective density change in the Gelsil due to helium adsorption is
\begin{equation}
 \frac{\delta\rho_\mathrm{g}}{\rho_\mathrm{g0}} = \frac{m nN_\mathrm{A}S}{m_\mathrm{g0}} \leq \frac{p\rho_\mathrm{liq}}{\rho_\mathrm{g0}},
 \label{eq:rho}
\end{equation}
where $m$ is mass of a helium atom, $n$ the coverage, $N_\mathrm{A}$ Avogadro's constant, and $\rho_\mathrm{liq}$ the density of bulk liquid helium. %修正181013
$S=166\ \mathrm{m^2}$, $m_\mathrm{g0}=0.371$ g, $p=0.54$, and $\rho_\mathrm{g0}=0.954\ \mathrm{g/cm^3}$ are the surface area, the mass, the porosity, and the density of the Gelsil sample, respectively. %修正181013
For example, $n=23\ \mathrm{\mu mol/m^2}$ of $^4$He film gives $\delta \rho_\mathrm{g}/\rho_\mathrm{g0}=0.041$ and $2\delta f/f_0 = -5.5\times 10^{-6}$.
%, which is an order of magnitude smaller than the observed values. 
Even if the pores are filled with liquid $^4$He, % (upper limit of weighting), 加圧できるのでupper limitではないのでは。%確かにそうですね．わかりました181016巻内
it gives $\delta \rho_\mathrm{g}/\rho_\mathrm{g0}=p\rho_\mathrm{liq}/\rho_\mathrm{g0}=0.08$, hence $2\delta f/f_0 = -1\times 10^{-5}$.
Therefore, if we measure a frequency increment greater than $2\delta f/f_0 \sim 1\times 10^{-5}$ ($\delta f \sim 4$ mHz) for any coverage of $^4$He or $^3$He, it is explained by changes in elasticity. Mass decoupling by the superfluid and ``supersolid'' transitions and by slippage phenomenon \cite{Hieda2000} is excluded from the origin. %修正181013。檜枝君のを引用しては%しました
%The absolute values of these frequency shifts are small and negligible compared to the measured frequency shift $2\delta f/f_0 \sim 1\times 10^{-4}$. 
%The measured frequency shift $2\delta f/f_0 \sim 1\times 10^{-4}$ is therefore explained by the change in elasticity, $\delta G_\mathrm{g}/G_\mathrm{g0}\sim 0.0005$.

 \subsection{Experimental procedure}
 %overview
 We first performed the measurement for $^4$He films. % at low temperatures.は当たり前では。181013
 Then the TO was warmed up to room temperature to get rid of $^4$He, and the measurement for $^3$He was made.
 For each run, the resonant frequency and energy dissipation of the TO without helium film were measured at first. %修正181013

 %TO velocity, strain
 The TO was forced to oscillate electrostatically at the resonant frequency $f$ using a loop circuit.
The driving was made by applying a pulsed voltage of 1.5 V and a width of 50 $\mathrm{\mu s}$, with dc bias voltage 200 V$_\mathrm{dc}$, which was applied to the dummy bob electrode. %修正181013
In this condition, the strain applied to Gelsil rod is estimated to be $1.6\times 10^{-7}$, and the maximal velocity at the Gelsil rim near the dummy bob is 15 $\mu$m/s. %ここの最大速度とは、ダミーボブのすぐ近くの部分ですね?(ロッドの根元は捻れないのでは)181013 %おっしゃるとおりです．near the dummy bobを加えました181015巻内
We confirmed that the oscillation amplitude is linear to the drive voltage around this condition.
%\textcolor{blue}{
The resonant frequency $f$ was measured by a frequency counter stabilized by a rubidium frequency standard. %} %追加しました181015巻内 Typoと追加181015%ありがとうございます181016巻内

 %Ring down (Q factor)
 The dissipation $Q^{-1}$ was taken from the current $R$ due to capacitive change measured by a lock-in amplifier. %修正181013 %currentに修正しました．181015巻内
 After the drive voltage is stopped, $R$ decreases exponentially with time; $R(t)=R_0e^{-t/t_0}$, where $t_0$ is a relaxation time. %修正181013
 The inverse of the dissipation gives Q factor of oscillation, which is $Q=\pi f t_0$. %修正181013
 The Q factor is proportional to the current at a drive, $Q=cR$, and the constant $c$ was measured before the warming and after the cooling at the lowest temperature for each coverage. %at a driveを追加．ドライブを切ったときは比例関係がないため181015巻内
 We confirmed that $c$ does not change during a run. %このrunとは4He、3He各々全体のrunですか?わかりにくいです181013 %修正181013
 For the $^4$He run, $c = 2.48 \times 10^{13}\ \mathrm{A}^{-1}$, and for the $^3$He run, $c = 2.06 \times 10^{13}\ \mathrm{A}^{-1}$.
The dissipation $Q^{-1}$ at each temperature is obtained from the corresponding current $R$.  %修正181013

\begin{figure}[tb]
 \begin{center}
  \centering
  \includegraphics[width=70mm]{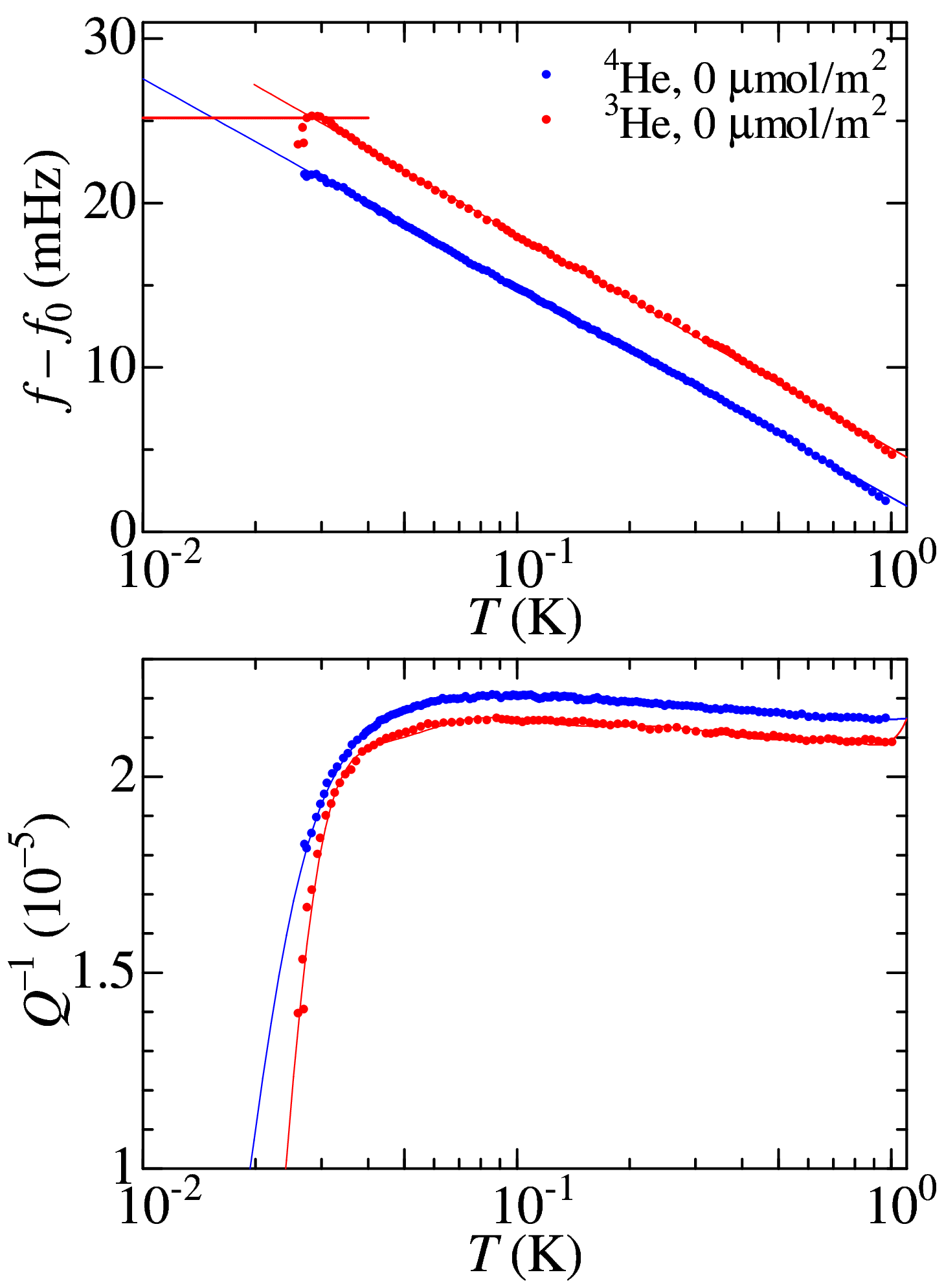}
  \caption{The background data of the empty cell.
  The upper panel is the resonant frequency $f$ from which a constant $f_0$ is subtracted.
  Here $f_0=860.82$ Hz for the $^4$He run, and $f_0=860.14$ Hz for the $^3$He run. 
 The lower panel is the dissipation $Q^{-1}$ for the $^4$He and $^3$He runs. Lines are results of the fitting (see text).}
  \label{fig:raw-fit}
 \end{center}
\end{figure}

 %resonant frequency
We refer to the temperature dependencies of $f$ and $Q^{-1}$ without helium ($n = 0$, empty cell) as the background. 
The resonant frequencies of the TO at $n = 0$ and at 1.0 K were $f_0=860.822$ Hz for $^4$He and $f_0=860.145$ Hz for $^3$He run.  %修正181013
The slight difference between two runs was by a thermal cycle.  %修正181013
%fitting of frequency
Fig. \ref{fig:raw-fit} shows the resonant frequency $f$ from which a constant $f_0$ is subtracted and the energy dissipation $Q^{-1}$ of $n=0$ for the $^4$He and $^3$He runs as a function of temperature. 
The frequency was found to be linear in $\log(T/\mathrm{K})$ at $T>20$ mK. 
We fit $f$ by 
\begin{equation}
 f(T,n=0) = \sum_{i=0}^1 A_i\left[\log(T/\mathrm{K})\right]^i.
\end{equation}
The fitting results are shown in Fig. \ref{fig:raw-fit}. 
For the $^3$He run, $f$ takes a maximum at about 30 mK and decrease with further lowering $T$, probably by the effect of tunneling two-level systems (TLS) in the glass sample\cite{EnssBook}.  %修正181013
We assume that $f$ is constant below 30 mK because the number of data are not sufficient to fit the $T$ dependence.

 %fitting of dissipation
 The dissipation $Q^{-1}$ slightly increases as $T$ decreases from 1 K to 80 mK, followed by a sudden drop below 50 mK. 
We fit $Q^{-1}$ by polynomial
\begin{equation}
 \log[Q(T,n=0)] = \sum_{i=0}^9 B_i\left[\log(T/\mathrm{K})\right]^i.
\end{equation}

 %Cooling
 For preparing an adsorbed helium film, a known amount of helium gas was admitted from a room temperature gas-handling system with 1 L standard volume to the TO at $T < 150$ mK.   %修正181013 
We have used commercial G1 grade $^4$He gas with impurity concentration less than $5\times 10^{-7}$, and $^3$He gas with nominal purity 99.95 \%.
After adsorbing helium gas at low temperature, the TO was warmed up at sufficiently high temperature, typically 1--5 K, for several hours to uniformly spread out in the Gelsil.  %修正181013
The TO was again cooled to 10--30 mK and then warmed up to 1.1 K to measure the temperature dependence of the resonant frequency $f$ and the dissipation $Q^{-1}$.
Data shown in this paper were taken during the warming.  %修正181013
The warming was done with PID control of the heater power while the dilution refrigerator was properly operated.  %修正181013
After the warming, the heater was turned off and the TO was cooled, and adsorption for the next coverage was started. 
No hysteresis was observed in the data between the warming and cooling.  %修正181013
 
 %Annealing
The annealing temperature and duration were selected so that the frequency and the amplitude become stable. 
For $^4$He films of coverage $n <15\ \mu$mol/m$^2$, the annealing was done at 5 K for 5 hours. 
At $16 \le n \le 26\ \mu$mol/m$^2$, it was done at 1.1 K for more than 14 hours. 
For $^3$He films, we annealed at 5 K for 5 hours for all coverages. 
For 16 $\mu$mol/m$^2$ of $^3$He, we first annealed at 1.1 K for 12 hours as in the case of $^4$He film. 
However, this condition was not sufficient because the frequency and dissipation were almost the same as those of previous $n=15\ \mu$mol/m$^2$ data. 
This indicates that $^3$He atoms in the extended state (see later section) are less mobile than $^4$He atoms. %ここはextended stateをまだ言っていないのでわかりにくいのでは.
%考えて下さい181013
We finally found that the annealing at 5 K for 5 hours was sufficient for $^3$He.

\section{\label{sec:results}Results and Discussion}

\subsection{Raw data}

\begin{figure}[tb]
 \begin{center}
  \centering
  \includegraphics[width=\columnwidth]{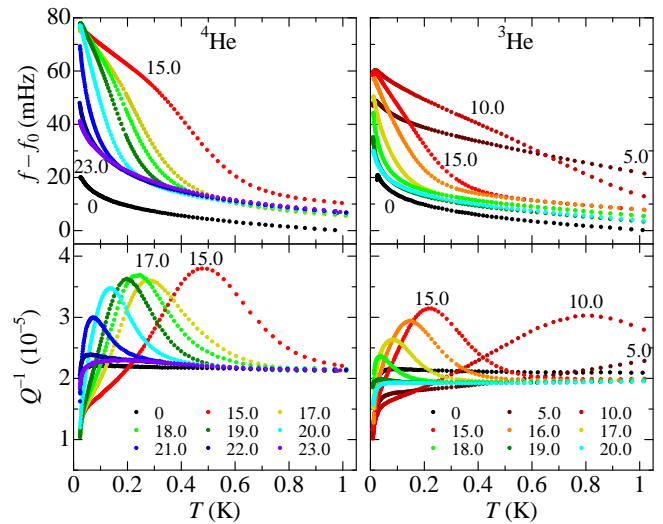}
  \caption{The resonant frequency $f$ and dissipation $Q^{-1}$ of the TO for $^4$He (left) and $^3$He films (right). 
Numbers give the coverage $n$ in unit of $\mathrm{\mu mol/m^2}$. 
Correspondence between $f$ and $Q^{-1}$ is shown by colors of data. 
All the $f$ data are shown after subtracting a constant frequency $f_0$, which is 860.822 Hz for $^4$He and 860.145 Hz for $^3$He, respectively.
  }
  \label{fig:raw}
 \end{center}
\end{figure}

In Fig. \ref{fig:raw}, we show raw data for the coverage $n$ from 0 to 23 $\mu$mol/m$^2$ for $^4$He, and to 20 $\mu$mol/m$^2$ for $^3$He, respectively. %(approximately 2 atomic layers)イントロの文から推測可能なため省略(巻内) 4Heの後ろにcommaを打ちました 180810巻内
%The data at $n = 0$, which is referred to as the background, are discussed in SM. %empty cellを削ってbackgroundではどうですか。SMも修正します
The dots in black are the background ($n = 0$). 
%Fig 1.のempty cellのデータをなくすという意味ですか？それともempty cellという表記をbackgroundと言い換えると言う意味ですか？(巻内)
%empty cell backgroundを単にbackgroundとしても通じるのではないか、という意味です。もしくは、ECBというabbreviationにしてもよい。
%承知しました(巻内)
Helium adsorption ($n>0$) increases $f$ in the entire temperature range from the background. 
%TO frequency is given by $f=(1/2\pi)\sqrt{k/I}$, where $k$ is a torsion constant (stiffness) of the rod and $I$ is a moment of inertia of the dummy bob.
%As $I$ is constant, % in our TO,省略(巻内)
%the increase in $f$ indicates that helium adsorption stiffens the rod.
%As the frequency increase is of the order of 10 mHz, it definitely shows that helium adsorption stiffens the torsion rod.

At each coverage, $f$ increases as $T$ decreases more rapidly than the background does. %下から持ってきました181014
By comparing the data with FEM simulations, we find that the observed increase in $f$ is originated from change in elasticity of helium adatoms. %修正181014
%We have computed the $\delta f$--$\delta G_\mathrm{g}$ and $\delta f$--$\delta\rho_\mathrm{g}$ relations by FEM simulations\cite{SM}. %executing略
%Linear relations $2\delta f/f_0 = 0.197\delta G_\mathrm{g}/G_\mathrm{g0}$ and $2\delta f/f_0 = -1.33\times 10^{-4} \delta \rho_\mathrm{g}/\rho_\mathrm{g0}$ are obtained. %省略(巻内) for changes in $G$ and $\rho$. %small略
The FEM simulation shows that if 15 $\mathrm{\mu mol/m^2}$ of $^4$He were decoupled from the oscillation, $f$ would increase by 1.5 mHz, % at the most, 略
which is nearly two orders of magnitude smaller than the observed increment at lowest $T$, $\delta f \sim 50$ mHz.  %修正181014 %50 mHzは例えばn=15の結果でしたので，比較するマスデカップリングによる周波数増加もその吸着量にしました181015巻内
Therefore, the increase in TO frequency is not due to the change in mass loading, i.e. superfluidity, supersolidity and slippage of helium films, %supersolidity略 %修正181014
but is originated from stiffening of helium films. %the stiffeningのtheを略(巻内)

%At each coverage, $f$ increases as $T$ decreases more rapidly than the background does.上に持って行きました181014
The dissipation $Q^{-1}$ has a peak at a temperature where the slope of $f$ is the largest, and its position decreases with increasing $n$. 
As $T$ decreases further, $f$ tends to saturate and $Q^{-1}$ decreases. %上から移動180723修正181014
These behaviors of $f(T)$ and $Q^{-1}(T)$ are qualitatively the same for $^4$He and $^3$He films. 
%\textcolor{blue}{% [Referee A (1)]
We call these phenomena \textit{elastic anomaly}. %, which is observed in both $^4$He and $^3$He films.ここは必要ですか?すぐ上で4He3He同じと書いてあります181014
The elastic anomaly vanishes at $n\simeq 23$ and 20 $\mathrm{\mu mol/m^2}$ for $^4$He and $^3$He, respectively.
 We can regard these coverages as the critical coverage $n_\mathrm{c}$. %いれました181014
We will discuss later that $n_\mathrm{c}$ determined from the elastic anomaly in $^4$He is %laterをいれprobably を取って181014
identical to $n_\mathrm{c}$ for the onset of superfluidity within the experimental accuracy. %最後にいれました181014
 %We confirmed that the data from $n_\mathrm{c}$ to %ここは実験事実の記述ですが、なぜconfirm(予測を確かめる)なのかわかりませんでした181014
At coverages $n > n_\mathrm{c}$, %25 and 40 $\mathrm{\mu mol/m^2}$ for $^4$He and $^3$He, respectively), 
both $f(T)$ and $Q^{-1}(T)$ are almost identical to those of the $n=0$ background, except for small upward shift in $f$ at all temperatures ($+6$ and $+3$ mHz for $^4$He and $^3$He, respectively). %修正181014
Therefore, in the superfluid phase of $^4$He and liquid phase of $^3$He, no prominent elastic anomaly is observed. 
It is remarkable that at $n > n_\mathrm{c}$ the TO behaves as if there were no adsorbed helium except the temperature-independent shift. %この文を付け足しましたが、下の段落で言いたいことがはっきりすれば必要ない気もします181014 %except the temperature-independent shiftを足しました181014巻内
% show  do not show much difference, which assures that films after superfluid or liquid appears have no elasticity effect to the same extent of the elastic anomaly.ここの意味がよくわかりませんでした181014
%}

The gradual increase in $f$ suggests a crossover of helium film from a soft to a stiff state, not a first order phase transition such as solidification. %1次相転移の文言を戻してみました。が、この一文はすぐ下のAnelasticmodelの節の頭に持って行っても良いのでは。181014
Comparing the resonant frequencies at $n=$ 15 and 23 $\mathrm{\mu mol/m^2}$ of $^4$He, for instance, we see that $f(15\ \mathrm{\mu mol/m^2})$ is larger than $f(23\ \mathrm{\mu mol/m^2})$ in the entire temperature range.
This means that the thinner film has a \textit{larger} elastic constant than the thicker film does. %constant181016
Such a coverage dependence of elastic anomaly can never be explained by the inert layer model. 
% is inadequate to explain such a coverage dependence of the elastic anomaly, and thus a model such that we discuss as below is essential.
We show below that a two-band model considering gapped excitation in the localized state explains qualitatively the observed elastic anomaly. %これでどうでしょうか。181016 %ありがとうございます181016巻内
%}
%%%%%%%%%%

\subsection{Anelastic model and energy gap}
 %Anelastic model
The temperature dependencies of $f$ and $Q^{-1}$ are typical of a relaxational crossover between a soft state at high $T$ and a stiff state at low $T$ under AC stress applied to a substrate-He system. % a system of substrate and He adatoms. を短縮(巻内)
Assuming that the relaxation is caused essentially by adsorbed helium, the relaxational contribution to $f$ and $Q^{-1}$ is obtained by subtraction of the background from the raw data.
 We define the frequency shift by
\begin{equation}
 \delta f \equiv f(T,n) -f(T,0)- [f(1\ \mathrm{K},n) - f(1\ \mathrm{K},0)].
\end{equation}
By this definition, we have omitted the background and the small constant frequency increments which were seen for all coverages (see $f$ at high temperatures), so as to set $\delta f = 0$ at 1.0 K.
%This omission was necessary for the fitting of data to the response function described in the main text, and did not give influence to the analysis described in the text, except that the 2D compressibility of helium film does not show divergent behavior in reality.この文は現在削除ですが、戻しても良いと思います181014%考えてみて下さい181014
%\textcolor{blue}{
The temperature-independent extra background is attributed to adsorption of atoms in particularly deep potential sites on the disordered substrate, and this omission was necessary for the fitting of data to the response function given later. %} %このように直して戻しました．2D compressibility に関するところは，この後のCに書きました181014巻内
%とりあえず良いです。181016
We also define the excess energy dissipation by
\begin{equation}
 \delta Q^{-1} \equiv Q^{-1}(T,n)-Q^{-1}(T,0).
\end{equation}
%\textcolor{blue}{
This definition was sufficient for the data of $^4$He, but we have added a small constant to set $\delta Q^{-1}(1\ \mathrm{K}) = 0$ for the data of $^3$He. %} %3Heのデータ(Appendix Fig. 9)に対してだけは定数を足していました．それを記述します．181016巻内

\begin{figure}[t]
 \begin{center}
  \centering
  \includegraphics[width=\columnwidth]{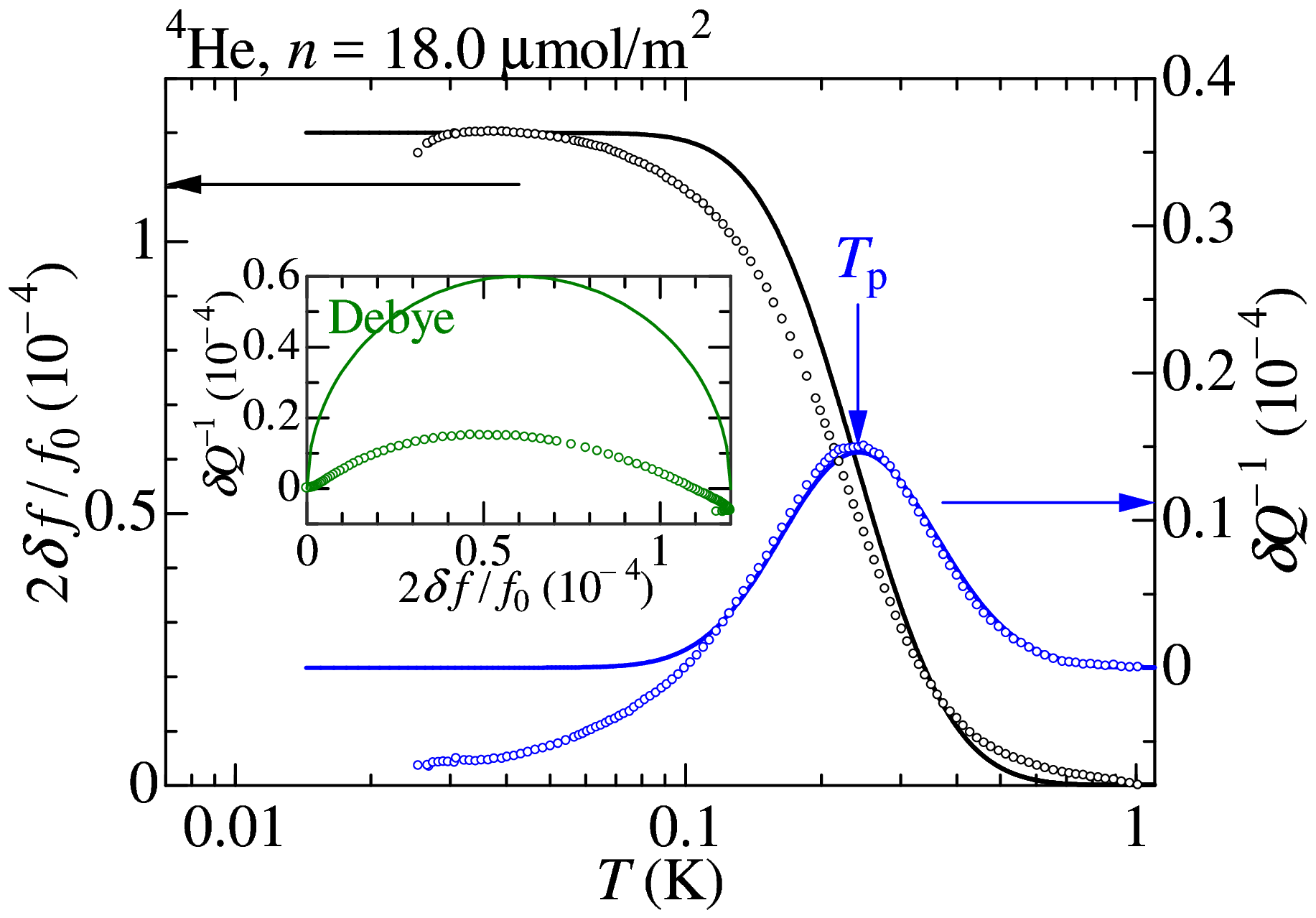}
  \caption{The normalized resonant frequency ${2\delta f}/{f_0}$ and excess dissipation $\delta Q^{-1}$ for a $^4$He film at $n = 18\ \mu$mol/m$^2$.
  Solid curves are the results of fitting to the response function, Eq. (\ref{eq:dfdQinv}). 
Fitting parameters are as follows: $\delta G/G_0 = 1.20 \times 10^{-4}$, $\tau_0 = 0.4\ {\mathrm{ns}}$, $\Delta/k_{\mathrm B} = 3.1\ {\mathrm{K}}$, and $\sigma = 0.38$. A vertical arrow shows $T_{\mathrm p}$. 
 Inset is Cole--Cole plot of the data. Semicircle shows the Debye relaxation with a single $\tau$. 
  }
  \label{fig:fitting}
 \end{center}
\end{figure}

Figure \ref{fig:fitting} shows a normalized frequency shift $2\delta f/f_0$ and an excess dissipation $\delta Q^{-1}$ for $^4$He at $n=18\ \mu$mol/m$^2$.
Other data, including $^3$He data, are presented in Figs. \ref{fig:data4He} and \ref{fig:data3He} in Appendix. %少し修正181014 
%where $\delta f$ is obtained by subtracting the background and a constant so as to set $\delta f = 0$ at 1.0 K, 
%and $\delta Q^{-1} = Q^{-1} - Q^{-1}_0$, where $Q^{-1}_0$ is the background.
%ここのコメントアウト行も戻すかどうか検討下さい181014 %ここのコメントアウトは，前の段落で詳細に述べていることなので，このままでいいと思います181014巻内
The dissipation $\delta Q^{-1}$ becomes negative below 0.1 K, meaning that helium adsorption decreases the internal loss of the glass. 
The physical origin of this apparently negative dissipation has not been elucidated. %意味ない文ですが、いれてみました。181014
%This is probably due to blocking of the motion in tunneling two-level systems in glass by helium adatoms\cite{EnssBook}. %ここまで言えるのか，疑問を持ってきました．ひとつ上の文だけ留めるのはどうでしょうか180811巻内
%TLSに言及できるのならしてください181014

\begin{figure}[tb]
 \begin{center}
  \centering
  \includegraphics[width=70mm]{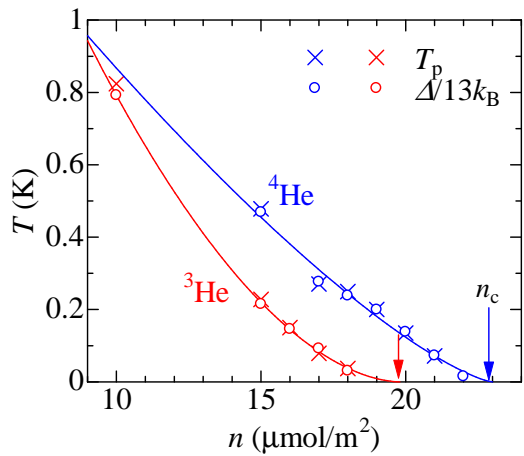}
  \caption{The dissipation-peak temperature $T_{\mathrm p}$, and the energy gap $\Delta$ obtained from the fittings as a function of the coverage.
  Solid curves are power low fits for $\Delta$ (see text).
  Arrows indicate critical coverage $n_{\mathrm c}$. %See text for the derivation of $\Delta$ and $n_{\mathrm c}$.これを削除
  }
  \label{fig:gap}
 \end{center}
\end{figure}

The dissipation-peak temperature $T_\mathrm{p}$ is indicated by an arrow in Fig. \ref{fig:fitting}. 
It can be recognized as a crossover temperature between the stiff and the soft state. %分けてみました181014 %文頭のTpをitに変えました181014巻内
The coverage dependence of $T_\mathrm{p}$ is shown in Fig. \ref{fig:gap}. %文頭はtheでは181014
 $T_\mathrm{p}$ approaches 0 K at $n=n_\mathrm{c}$ with a concave curvature. %修正、smoothlyを取ってみました181014

The relaxational crossover is explained by %the two-band は後と重複しているので省略してみました180812
a thermal activation process of helium adatoms between two discrete energy bands\cite{Tait1979,Crowell1995}. %aを付加180812
At $T=0$, helium atoms are localized and forms an energy band. 
At finite temperatures, the localized atoms are thermally excited to another band of extended states separated by an energy gap.
%The excited atoms move freely on the substrate. %(normal fluid). %(normal fluid)追加180810巻内180812変更
%ここに(normal fluid)とあっても意味が判らないのでは。The excited atoms move freely on the substrate, and act as a normal fluid.などでは。180812
The excited atoms move freely along the substrate and act as a normal fluid. %Fig. 4にあるnormal fluidの説明が本文にないといけないと思い，先生の案に替えました180812巻内 %on the surface -> along the surface 181010巻内
We analyze $2\delta f/f_0$ and $\delta Q^{-1}$ by dynamic response functions for anelastic relaxation, according to similar anelasticity analysis for bulk solids\cite{NowickBook} %付け足してみました181014
\begin{eqnarray}
 &&\frac{2\delta f(T)}{f_0} = \frac{\delta G}{G_0} \left[ 1 -  \frac{1}{1+[\omega \tau(T)]^2} \right], \\
 &&\delta Q^{-1}(T) = \frac{\delta G}{G_0} \frac{\omega \tau(T)}{1+[\omega \tau(T)]^2},
\end{eqnarray}
where $\delta G$ and $G_0$ are a relaxed shear modulus and a shear modulus of TO rod respectively, and $\omega = 2\pi f \simeq 2\pi f_0$.
%\textcolor{blue}{
The dissipation $\delta Q^{-1}$ has a peak at $\omega \tau=1$. %} %追加181014巻内
The thermal relaxation time is given by %$\tau(T)$省略 180810巻内
$\tau(T) = \tau_0 e^{E/k_\mathrm{B}T}$, 
where $E$ is an energy gap (an activation energy) and $\tau_0^{-1}$ is an attempt frequency. %$\tau_0$ an inverse attempt frequencyから変更180811巻内
%\textcolor{blue}{
Now it becomes clear that the dissipation-peak temperature $T_\mathrm{p}$ is the temperature which holds $1 = \omega\tau_0 e^{E/k_\mathrm{B}T_\mathrm{p}}$. %} %追加しました．また，次の文を段落分けしました181014巻内

If $\tau(T)$ were single valued, the relaxation would be a Debye type and the plot of $2\delta f/f_0$ versus $\delta Q^{-1}$ would be a semicircle shown in the inset of Fig. \ref{fig:fitting}. %shown付加181014
The plot is, however, a deformed semicircle, meaning that $E$ has a distribution.

We assume a log-normal distribution for $E$ %修正181014
\begin{equation}
 F(E) = \frac{1}{\sqrt{2\pi}\sigma E} \exp \left(-\frac{[\ln (E/\Delta)]^2}{2\sigma^2}\right),
\end{equation}
where $\Delta$ is the median %\textcolor{blue}{
(a value separating the higher half of the population from the lower half). %}.
We hereafter regard $\Delta$ as the energy gap.
The selection of a log-normal distribution is reasonable because $\delta Q^{-1}(T)$ is almost symmetric for $\log(T)$ scale as shown in Fig. \ref{fig:fitting}.
We obtain a complex form of the dynamic response function as 
\begin{equation}
\label{eq:dfdQinv}
 \frac{2\delta f}{f_0} +i\delta Q^{-1} = \frac{\delta G}{G_0} \left[ 1 -  \int_0^\infty \frac{F(E)}{1+i\omega \tau(E,T)} dE \right]. 
\end{equation}

%BG subtraction and fitting
We perform fittings of Eq. (\ref{eq:dfdQinv}) to the data.
% The background-subtracted data of four coverages of $^4$He and $^3$He are shown with the results of fitting in Fig. \ref{fig:data4He} and \ref{fig:data3He}, respectively. 
%For both $^4$He and $^3$He films, the normalized resonant frequency shift $2\delta f/f_0$ increases at lower temperature accompanying a dissipation peak.
The results are shown with solid curves in Fig. \ref{fig:fitting}, and Figs. \ref{fig:data4He} and \ref{fig:data3He} in Appendix. %Figs.とAppendixの順番を入れ替えました181015巻内
Equation (\ref{eq:dfdQinv}) fits well to the data, and the negative $\delta Q^{-1}$ below 0.1 K does not give much influence to the quality of the fittings. %文頭がEq.で始まっていたため，修正． apparently省略 180810巻内

We find a remarkable relation between $\Delta$ and $T_{\mathrm p}$, $\Delta \simeq 13k_{\mathrm B}T_{\mathrm p}$, for both $^4$He and $^3$He, as clearly shown in Fig. \ref{fig:gap}. %強めてみました181014
This relation between $\Delta$ and $T_{\mathrm p}$ reinforces the validity of the fittings. %修正181014
 Other fitting parameters, $\delta G/G_0$, $\tau_0$ and $\sigma$, have no systematic dependencies on $n$. 
 The inverse attempt frequency $\tau_0$ is ranged from 0.4 to 0.6 ns, %\textcolor{blue}{
 which holds the aforementioned relation $1 = \omega \tau \simeq \omega \tau_0 e^{13}$. %} %1=ωτでTpということを説明していない．省略(巻内)%了解です%これは戻しても良いのではないでしょうか。検討ください181014 %戻しました．これに伴い，1=ωτで散逸にピークがあるということも前に記述しました181014巻内
%了解です181016

 %Energy gap: power law
Figure \ref{fig:gap} shows that $\Delta$ and $T_{\mathrm p}$ monotonically decrease with some scatters. %scattersを戻しました181014
%を省略180811巻内
The gap is fitted by a power law %$\Delta$ -> The gap 180810巻内
\begin{equation}
 \label{eq:delta}
 \Delta = \Delta_0|1 - n/n_\mathrm{c}|^a.
\end{equation}
 Nonlinear fittings give $\Delta_0/k_\mathrm{B}=23.9\ \mathrm{K}$, $n_\mathrm{c} = 23.0\ \mathrm{\mu mol/m^2}$ and $a = 1.32 $ for $^4$He, and $36.5\ \mathrm{K}$, $ 19.8\ \mathrm{\mu mol/m^2}$ and $ 1.80 $ for $^3$He, respectively.

 \subsection{Energy band and compressibility}

\begin{figure}[tb]
 \begin{center}
  \centering
  \includegraphics[width=80mm]{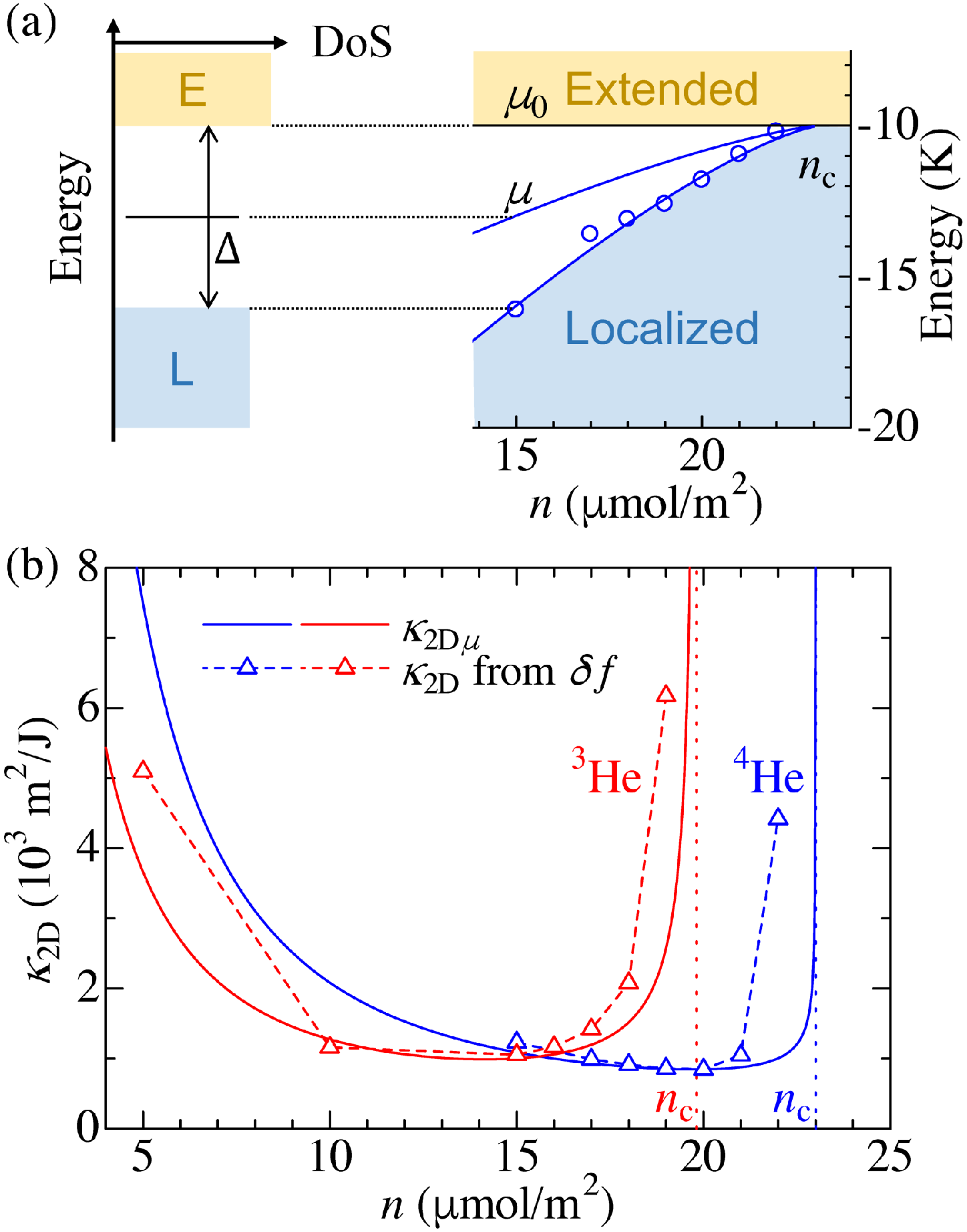}
  \caption{(a) Proposed energy band for helium films.
  The left side shows density of states (DoS) of the localized and extended states separated by a gap $\Delta$.
  The chemical potential $\mu$ is located at the middle of the gap. % at 0 Kを省略(巻内)
  The right side shows its $n$ dependence.
  The uppermost energy of the localized states increases with $n$, while the lowermost energy of the extended states stays at $-$10 K, which was obtained as the chemical potential of $^4$He film on glass from a numerical study\cite{Boninsegni2010}. %LとEの略は必要ないと思います。InsetでもLocalized, extended で見えるのでは? %文章中のLとEの略はやめました．(巻内)
  (b) 2D compressibility of $^4$He and $^3$He films.
  Solid curves show $\kappa_{\mathrm{2D\mu}}$ from Eq. (\ref{eq:kappa}) and triangles show $\kappa_{\mathrm{2D}}$ directly obtained from $\delta f$ using Eq. (\ref{eq:kappa2d}).
  %Inset shows the definition of $\delta f'$.ビジーだったのでこのインセットは無しにしました(巻内)
  }
  \label{fig:band}
 \end{center}
\end{figure}

 The fact that $\Delta$ smoothly decreases to zero as $n \rightarrow n_{\mathrm c}$ indicates that the energy band also smoothly changes with $n$.
 We propose an energy band in Fig. \ref{fig:band}(a). %前の文から独立180811巻内
The localized states are completely filled at $T=0$, and its uppermost edge is determined by $n$.
Atoms in the localized states contribute to the elasticity.
On the other hand, the extended states are empty at $T=0$, and their lowermost edge, $\mu_0$, has no or negligible dependence on $n$.
At high $T$, helium atoms are thermally excited from the localized to the extended states, resulting a softening.
At $n \ge n_{\mathrm c}$, the gap is closed, and helium atoms can enter the extended states even at $T=0$.
$^4$He atoms condensed in the extended states show superfluidity. 
This scenario was first discussed by Crowell \textit{et al.} in a heat capacity study of $^4$He films\cite{Crowell1995,*Crowell1997}. %to explain heat capacity -> in a heat capacity study 180810巻内

The band for each $n$ is analogous to that of intrinsic semiconductor.
The chemical potential $\mu$ at $T=0$ is a function of $n$ and is located at the middle of the gap, so that %二文に分けました．またshould be -> is 180810巻内
\begin{equation}
 \mu (n) = \mu_0 - \Delta(n)/2. %\frac{\Delta(n)}{2}. こうすることで縦幅が狭くなるので，文字数削減できるかもしれません180811巻内了解です。180812
  \label{eq:mu}
\end{equation}
%Shown in the inset of Fig. \ref{fig:delta} is $\mu(n)$ in the band. %少し省略 数行上でも述べているので，省略(巻内)
%$\Delta(n)$ obtained from the fittings is scattered, but one may assume that $\mu(n)$ changes smoothly with $n$. これもΔに関しては似た記述が数行上にあるので省略(巻内)
%Then, the derivative $d\mu/dn$ is related to the 2D compressibility as %dと\partialの使い分けていますか？結局，下のように短縮しました(巻内)
The 2D compressibility is, by definition, %了解です
%\begin{equation}
% \kappa_{\mathrm{2D\mu}} \equiv \left( \tilde{n}^2 \frac{\partial \mu}{\partial \tilde{n}} \right)^{-1} = \left( N_\mathrm{A} n^2 \frac{\partial \mu}{\partial n} \right)^{-1},
%\label{eq:2Dcompressibility}
%\end{equation}
%where $\tilde{n} =N_\mathrm{A}n$ is number density of atoms and $N_\mathrm{A}$ the Avogadro constant.
\begin{equation}
 \kappa_{\mathrm{2D\mu}} = \left( N_\mathrm{A} n^2 \frac{\partial \mu}{\partial n} \right)^{-1},
 \label{eq:2Dcompressibility}
\end{equation}
where $N_\mathrm{A}$ is the Avogadro constant ($N_\mathrm{A}n$ is a 2D number density of atoms).
We refer to the 2D compressibility obtained from $\mu(n)$ as $\kappa_{\mathrm{2D\mu}}$. 
From Eqs. (\ref{eq:delta}), (\ref{eq:mu}) and (\ref{eq:2Dcompressibility}), we get
\begin{equation}
 \kappa_{\mathrm{2D\mu}}  = \frac{2n_\mathrm{c}}{aN_\mathrm{A}n^2\Delta_0}|1-n/n_\mathrm{c}|^{1-a}. 
\label{eq:kappa}
\end{equation}
The results are drawn in Fig. \ref{fig:band}(b) with solid curves.

The 2D compressibility is also obtained by directly comparing the observed frequency increment $\delta f(n)$ with Eq. (\ref{eq:FEMg}), the FEM result. %, $2\delta f/f_0  = 0.197\delta G_\mathrm{g}/G_{\mathrm{g0}}$. SMからもってきたので，式番号ありました181015巻内
Here $\delta f(n)$ is frequency increment from $f(n_\mathrm{c})$ at the lowest temperature.
%This omission was necessary for the fitting of data to the response function described in the main text, and did not give influence to the analysis described in the text, except that the 2D compressibility of helium film does not show divergent behavior in reality.この文は現在削除ですが、戻しても良いと思います181014 %Bから．下のようにしてここに入れました181014巻内
%\textcolor{blue}{
We use $f(n_\mathrm{c})$, not $f(n=0)$, as the reference value because $f(n_\mathrm{c})$ contains the elastic contribution from atoms in deep potential sites which we want to exclude from the calculation. %} %追加しました．Bにおける上のコメントアウトした文を反映181014巻内
%了解です181016
The shear modulus of Gelsil is $G_{\mathrm{g0}} = 7.38$ GPa from an ultrasound study\cite{Negishi}. %引用にしました181014 %our->an 181015巻内
%少し変えました(巻内)
%Elastic modulus of He film is directly obtained by comparing the observed $\delta f$ with calculation of $f$ as a function of change in elastic constant of the rod. %少し省略
%The FEM calculation results in $2\delta f/f_0  = 0.197\delta G_\mathrm{g}/G_{\mathrm{g0}}$, where $G_{\mathrm{g0}} = 7.38$ GPa obtained from our ultrasound study (see SM).
%Elastic modulus of He film is directly obtained by comparing the observed $\delta f$ with the FEM result $2\delta f/f_0  = 0.197\delta G_\mathrm{g}/G_{\mathrm{g0}}$, where $G_{\mathrm{g0}} = 7.38$ GPa obtained from our ultrasound study (see SM). %上の二文を一文にまとめました．FEMの式は前出のため(巻内)
%Here we define the frequency increment $\delta f'(n) = f(T_\mathrm{min},n) - f(T_\mathrm{min},n\agt n_\mathrm{c})$, which is irrevant to the porposed band model (for further reasons, see SM).
%With a general relation between $G$ and Young's modulus $K$, $K = \lambda + (2/3)G$ where $\lambda$ is L\'ame constant, %for homogeneous and isotropic material 略
With a general relation $K = \lambda + (2/3)G$, where $K$ is a bulk modulus and $\lambda$ is a L\'ame constant, %書き換えて2 word削減(巻内)
an effective 3D compressibility of helium film $\kappa$ is $\kappa^{-1} = \delta K \simeq (2/3)\delta G$. %obtained as省略(巻内)
It is converted to the 2D compressibility by $\kappa_\mathrm{2D} =\kappa/d$, where $d = v_{\mathrm {film}}n$ is mean film thickness and $v_{\mathrm {film}}$ is molar volume of helium film. 
Since $v_{\mathrm {film}}$ is unknown, we employ $v$ of liquid helium at 0 bar, which might be larger than $v_{\mathrm {film}}$. %SVP→0 bar
Combining these equations, we have
 \begin{equation}
 \kappa_\mathrm{2D} = \frac{0.148 f_0}{\delta f(n)G_\mathrm{g0} v n}.
 \label{eq:kappa2d}
 \end{equation}
 In Fig. \ref{fig:band}(b), we plot $\kappa_{\mathrm {2D}}$ obtained from Eq. (\ref{eq:kappa2d}). 
 The overall agreement between $\kappa_{\mathrm{2D\mu}}$ and $\kappa_{\mathrm{2D}}$ from $\delta f$ definitely assures the proposed band.
 In both $^4$He and $^3$He, $\kappa_{\mathrm {2D}}$ first decreases, then makes a plateau, and finally shows divergent behavior as $n$ approaches $n_{\mathrm c}$. %thenいれました181014

 \subsection{Phase diagram} 
\begin{figure}[tb]
 \begin{center}
  \centering
  \includegraphics[width=80mm]{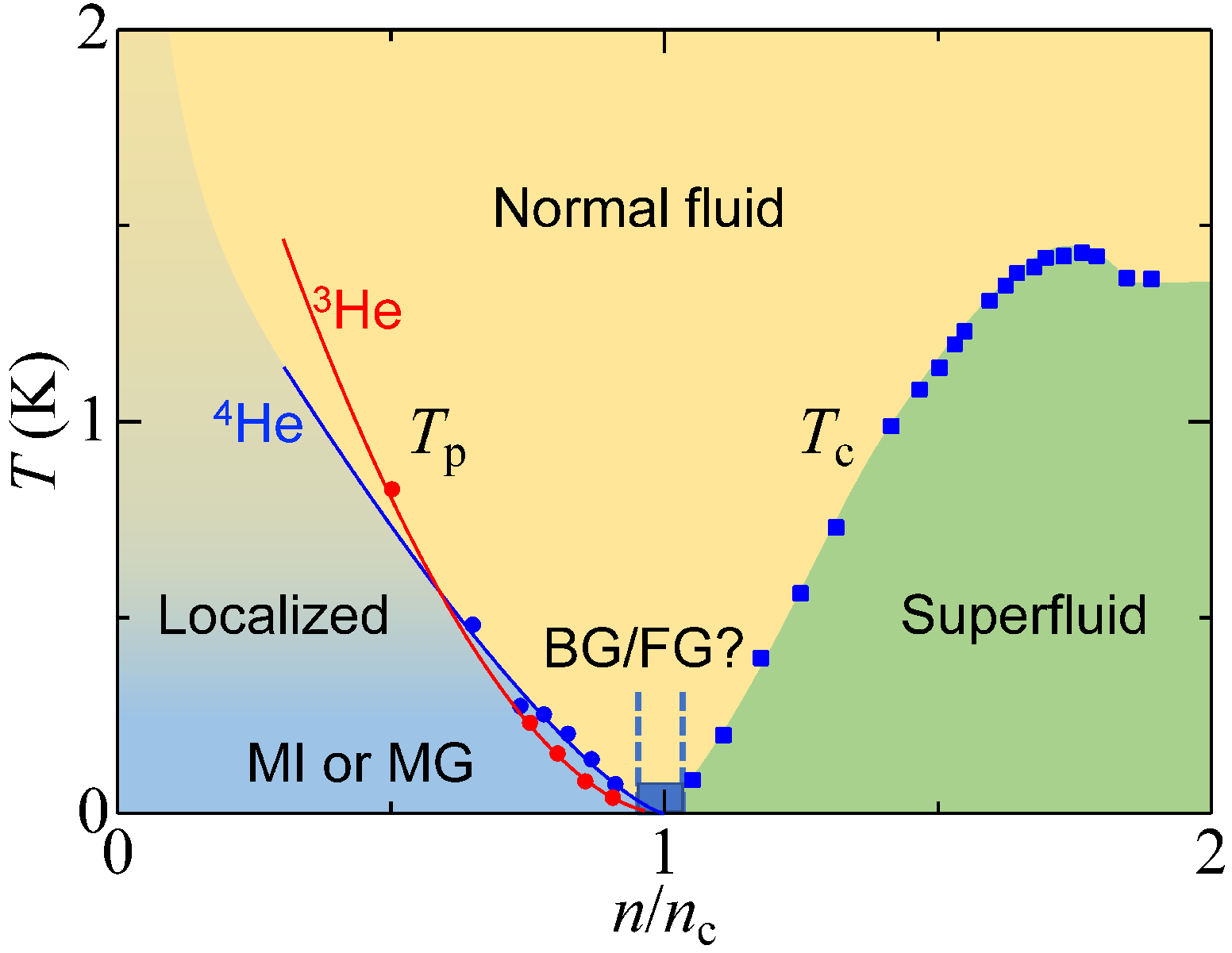}
  \caption{
  %\textcolor{blue}{(FIGURE MODIFIED. The blue region is extended to higher temperature.)}
  A unified phase diagram constructed by crossover temperature $T_{\mathrm p}$ and superfluid transition temperature $T_{\mathrm c}$ (only for $^4$He) as a function of $n/n_{\mathrm c}$. 
  The superfluid transition temperature is from previous TO study\cite{Yamamoto2004} %\textcolor{magenta}{
  ($^4$He on Gelsil, $n_\mathrm{c}=20\ \mathrm{\mu mol/m^2}$). %}. 
  The peak temperature divides the localized phase from normal fluid phase.
  The localized phase near $T=0$ is %\textcolor{magenta}{(self-organized$\rightarrow$)
  a sort of %}
  Mott insulator (MI) or Mott glass (MG).
  Possible region for Bose glass (BG) or Fermi glass (FG) is shown.
  }
  \label{fig:phasediagram}
 \end{center}
\end{figure}

 The universality in $^4$He and $^3$He films is revealed by constructing a ``unified'' phase diagram shown in Fig. \ref{fig:phasediagram}. %universalityにしてみましたがどうでしょうか181014 %よりいいと思います181014巻内
 The peak temperatures $T_{\mathrm p}$'s of $^4$He and $^3$He as a function of $n/n_{\mathrm c}$ nearly collapse onto each other, except that the curvatures differ. %文頭なので The peak temperatures を追加180810巻内
 %\textcolor{blue}{
 The difference of the zero-point energy, hence the binding energy from the substrate, between $^4$He and $^3$He does not affect the magnitude of the characteristic temperatures of the elastic anomaly. %} %181013修正しました巻内
%了解です181014

%この段落全体は、レフェリーのコメントを読んでいない人には唐突でわからないのでは。とても不自然な感じです。先のAのコメントを想定した段落もそうです181014
%\textcolor{magenta}{ [Referee B (1)]
Superfluid transition temperatures $T_\mathrm{c}$'s of $^4$He on Gelsil in a previous TO study\cite{Yamamoto2004} are also plotted in Fig. \ref{fig:phasediagram}.
The critical coverage for it was inferred to be $n_\mathrm{c} = 20\ \mathrm{\mu mol/m^2}$, which is slightly smaller than 23 $\mathrm{\mu mol/m^2}$ for the elastic anomaly. %修正181014 %the superfluidityだと，一般的にどの実験からもnc = 20となる，と誤解されると思いました．代名詞itに変えました181015巻内 
The nominal pore diameters were 2.5 nm for both porous Gelsil samples, but they were provided by different manufacturers. %typo181014
The discrepancy in $n_\mathrm{c}$ might be originated from differences in some characteristics such as residual impurities, pore size and its distribution between two samples. %residual impuritesを追加181015巻内
%了解です181016
%In a heat capacity study by Crowell \textit{et al.}\cite{Crowell1997}, they observed a crossover temperature for the gap for $n<n_\mathrm{c}$ and a superfluid transition temperature for $n>n_\mathrm{c}$ for one Vycor substrate. %これは本当ですか。熱容量とTOの測定は違うサンプルですよ181014 %熱容量(Vycor)から得たTcです．TO (aerogel)とは違います．Crowellは後ろで少し言及するにとどめました181015巻内
%全く同一の試料ではないと思いますが。同じロッドから切り出しても、結局熱処理等違えばncが違うこともあり得るので、Vyycorとの比較は必要か再検討ください181014
%Crowellの熱容量で出したギャップはnに対し直線的変化をしていたはずですね。またクロスオーバー温度なるものの外挿が超流動Tcの外挿と一致する物理的理由は?それがわからない場合比較の意味がないのでは181014
%The figure in their paper shows that these temperatures meet at the same critical coverage $n_\mathrm{c}=27\ \mathrm{\mu mol/m^2}$, within a few tenth of 1 $\mathrm{\mu mol/m^2}$. %結局ある精度内でしかncが同じと言えないので、ここの議論は意味ない気がしますが。181014 %承知しました．Crowellは後ろで少し言及するにとどめました181015巻内
%Another evidence is our previous TO study in Appendix B. %「証拠」とはどういう意味でしょうか。全く理解不能だと思います181014 %確かにevidenceではないですね181015巻内
%}

%\textcolor{magenta}{
In a previous TO study, %前の文を削除して，文頭につけました181015巻内
a TO with a Gelsil in the bob (named TO1, see Appendix B) %, which have a Gelsil disk (the nominal pore diameter is 2.5 nm) in the bob, detected an anomalous increase in the frequency with a dissipation peak, which now we identify with the elastic anomaly. %see Appendix Bを追加181016巻内
has detected both the superfluid transition and the elastic anomaly, the latter is confirmed by the present study.
The two characteristic temperatures, $T_\mathrm{p}$ and $T_\mathrm{c}$, % below and above $n_\mathrm{c}$
meet at the same critical coverage $n_\mathrm{c}=22\ \mathrm{\mu mol/m^2}$ in the experimental resolution (see Fig. \ref{fig:phasediagramKogure} in Appendix B). %TO1の測定(Fig13)でも、厳しく言うとTpとTcの外挿点が一致するとは言えないのでは。結局同一バッチの試料で100mK以下の測定もきちんとやって初めてわかることなので。181014 %実験精度内では，ということにとどめ，さらなる実験が必要ということを下で述べました．181015巻内
%As the amount of admitted helium gas is highly controlled, the discrepancy in $n_\mathrm{c}$ in several studies is attributed to the surface characterization of the substrates (roughness, impurities, and pore radius). %
A heat capacity study by Crowell \textit{et al.}\cite{Crowell1997} may also suggests the common critical coverage for ``$T_\mathrm{B}$'' and $T_\mathrm{c}$, though the physical meaning of $T_\mathrm{B}$, a crossover temperature of the heat capacity at $n<n_\mathrm{c}$, is not clear. %足しました181015巻内
Further experimental studies in the vicinity of $n_\mathrm{c}$ is necessary to conclude that the $n_\mathrm{c}$'s of the superfluidity and the elastic anomaly are exactly identical or slightly different.
%}
%%以上の段落も内容を考え直して下さい。ncの不一致は表面積測定や試料の問題で、TO1では測定精度内でncが一致するように見えると述べればよいでしょうか181014
%%上記のように書き換えました．いかがでしょうか？181015巻内
%さしあたり良いです。これで投稿して様子をみては181016
%承知しました181016巻内

Figures \ref{fig:gap} and \ref{fig:phasediagram} show that $\Delta(n)$ and $T_\mathrm{p}$ obey a power law $\Delta \propto T_\mathrm{p} \propto |n - n_{\mathrm c}|^a$ with $a > 1$.
A symmetry may exist between the critical exponent of $\Delta$ and that of superfluid $T_{\mathrm c}$ of $^4$He films, in which $T_{\mathrm c} \propto (n - n_{\mathrm c})^w$ with $w > 1$ in all previous results\cite{Csathy2003, Crowell1997}. % 指数をb -> w，慣習に習って180810巻内exists修正180812

Our finding is that $^4$He and $^3$He films at $n < n_{\mathrm c}$ are identically \textit{gapped} and \textit{compressible} irrespectively of quantum statistics. %文字数削減の変更180723巻内
These features do not strictly agree with the properties of Bose glass (gapped, compressible, for $^4$He)\cite{Fisher1989}, Mott insulator (gapped, incompressible) or Mott glass (single-particle gap, incompressible)\cite{Giamarchi2001}.
%MG has intermediate properties between MI and BG. %イントロで言っているので省略180810巻内
%字数制限が無くなったので、BGやMIなどを全部略さないで記しても良いと思います。検討ください181014 %略さないようにしました181014巻内

We propose, however, that the localized helium film is a sort of Mott insulator or Mott glass in a realistic situation. %aにしました181014 %films are->film is a sort of
One may consider the following %toy 要らないのでは181014 %分かりました．色もなくしました181015巻内
model: 
Helium atoms are first adsorbed on some particularly deep adsorption sites, so as to weaken randomness. 
Additional helium atoms are adsorbed on the weakened potential surface, and self-organize a nearly spatially periodic 2D Mott insulator or Mott glass with an $n$-dependent lattice %\textcolor{magenta}{(constant$\rightarrow$)
 spacing. %}. 
The self-organization of sites allows a finite compressibility. 
%\textcolor{magenta}{ [Referee B (2)]
The gap is finite because ``sites are fully occupied'' and an atom needs a finite energy to move.

Tackling this problem is important because it is related the nature of the onset of superfluidity, the quantum critical phenomena and the boson and fermion localization. %Tacklingなど多少修正181015巻内
%了解です181016 %改行しました181016巻内
Theoretical, numerical, and more experimental works are desired. %}

The gapped localized state which terminates at a certain coverage ($n_{\mathrm c}$) has been observed in helium films on various substrates, such as Vycor\cite{Tait1979,Crowell1997}, Hectorite (2D flat substrate), FSM (1D pores), and zeolites\cite{Wada2009}.
This suggests that the gapped localized Mott insulator or Mott glass ubiquitously exists, irrespectively of substrate randomness and dimensionality.
%この段落の文章の順番を変え，文を多少変えました180723
%MI, BGなどは省略しない方がインパクトありますが字数足りないですか?180812
%多用しているため，字数削減(13 words？)のため，あとFig. 4で省略を使っているため，このようにしました．180812巻内
%MI, BGなどの省略をやめました（Fig. 8内は除く）181014巻内

%(前から移動し，新たな段落に)
As to the $^4$He films, our result does not reject possibility of Bose glass in the vicinity of $n=n_{\mathrm c}$, where the gap is almost closed and the compressibility significantly increases. %compressibilityの記述を追加180723修正181014
Theories predict Bose glass emerging between Mott insulator and superfluid in the presence of moderate disorder\cite{Fisher1989}. %emergingといれました181014
The previous experiment discussed a quantum critical behavior of possible Bose glass near $n_\mathrm{c}$\cite{Crowell1997}. %criticalのtypo修正8/10
%変えてみました。がBGはどのくらい正当性あるのでしょうか180812
In our system, Bose glass can exist at about $22 < n < 23\ \mu$mol/m$^2$, and a corresponding Fermi glass can occur in $^3$He at $19 < n < 20\ \mu$mol/m$^2$. 
Recently, QPTs among Mott insulator, Mott glass, Bose glass and Bose-Einstein condensate are realized in a %bosonic 
quantum magnet\cite{Yu2012}. %areに修正181014
%Bosonic quantum magnetというのは確立した用語ですか?bosonicの意味はわかりますか?180812
%bosonicは自分で付けたので，確立した用語ではないと思います．S = 1です．この後でヘリウムではquantum statisticsを変えられる利点があると言うので，この系はboson系に限るということを示したかったのですが，無しでいいです．180812巻内
%Elusive localized phases become increasingly under experimental survey. %この文は要りますか?磁性でも局在相が重要という意味ですか181014 %コメントアウトしました．
Helium films in disordered substrates can open a new perspective of QPTs for advantage of variable correlation and quantum statistics. % (by choice of $^4$He or $^3$He). %choiceですか?一応修正8/10
%changeableよりvariableでは。またcoverageは不要です180812
%最後の括弧も、ここまで読んだらもうわかるので不要では。180812

\section{Conclusions}
%結論にもう少し加えても良い気がします。検討ください181014
%例えば水素やグラファイトの話など181014
%レフェリーからの新たな批判的，疑問的コメントを避けるため，新しいことは言及しなくても良いと思うのですが，いかがでしょうか181015巻内
%そうですね。それで良いです181016

We have discovered that the localized $^4$He and $^3$He films on a porous glass show an identical elastic anomaly. %anにしてみました181014　anomalies -> an anomaly 180810巻内 %on a porous glass追加181015巻内
The elastic anomaly is explained by thermal activation of helium atoms from the localized to extended states with a distributed energy gap, %anomalies are -> anomaly is 180810巻内
which decreases as the film approaches the critical coverage $n_{\mathrm c}$.
%\textcolor{blue}{
The two-dimensional compressibility showed divergent behavior near $n_\mathrm{c}$, which was deduced from the power low behavior of the gap and the energy band.
The divergent behavior of the compressibility was confirmed from the direct calculation of the observed frequencies. %} %追加しました181015巻内
Both the localized $^4$He and $^3$He are gapped and compressible, suggesting that the ground state is a sort of Mott insulator or Mott glass. % a 省略180810巻内 %self-organized->a sort of 181015巻内
Future studies in the vicinity of $n_{\mathrm c}$ at lower temperatures %and with other substrates such as graphite ここでグラファイトを出すのはイントロを考えると必要ないかと思い略しました。8/10 %承知しました180810巻内
will unveil the nature of the QPT.

\begin{acknowledgments}
 
 %小暮、吉村、東野君もTO1の実験で謝辞181014
 %\textcolor{blue}{
 We thank M. Kobayashi, T. Ohtsuki and A. J. Beekman for useful discussions, and T. Kogure, H. Yoshimura, R. Higashino and Y. Shibayama for the previous torsional oscillator study using TO1 in Appendix. %} %追加しました181014巻内
%良いです。一文にしてみました181016
 This work was supported by JSPS KAKENHI Grant Number JP17H02925.
 %\textcolor{blue}{
 TM was supported by Grant-in-Aid for JSPS Research Fellow 18J13209, Research Grant of Keio Leading-edge Laboratory of Science and Technology, and Keio University Doctorate Student Grant-in-Aid Program. %}
\end{acknowledgments}

\appendix{

\section{Additional data for the fittings}
We show in Figs. \ref{fig:data4He} and \ref{fig:data3He} additional data from which the background is subtracted.
The results of fittings to Eq. (\ref{eq:dfdQinv}) are also shown.

 \begin{figure}[tbh]
 \begin{center}
  \centering
  \includegraphics[width=\columnwidth]{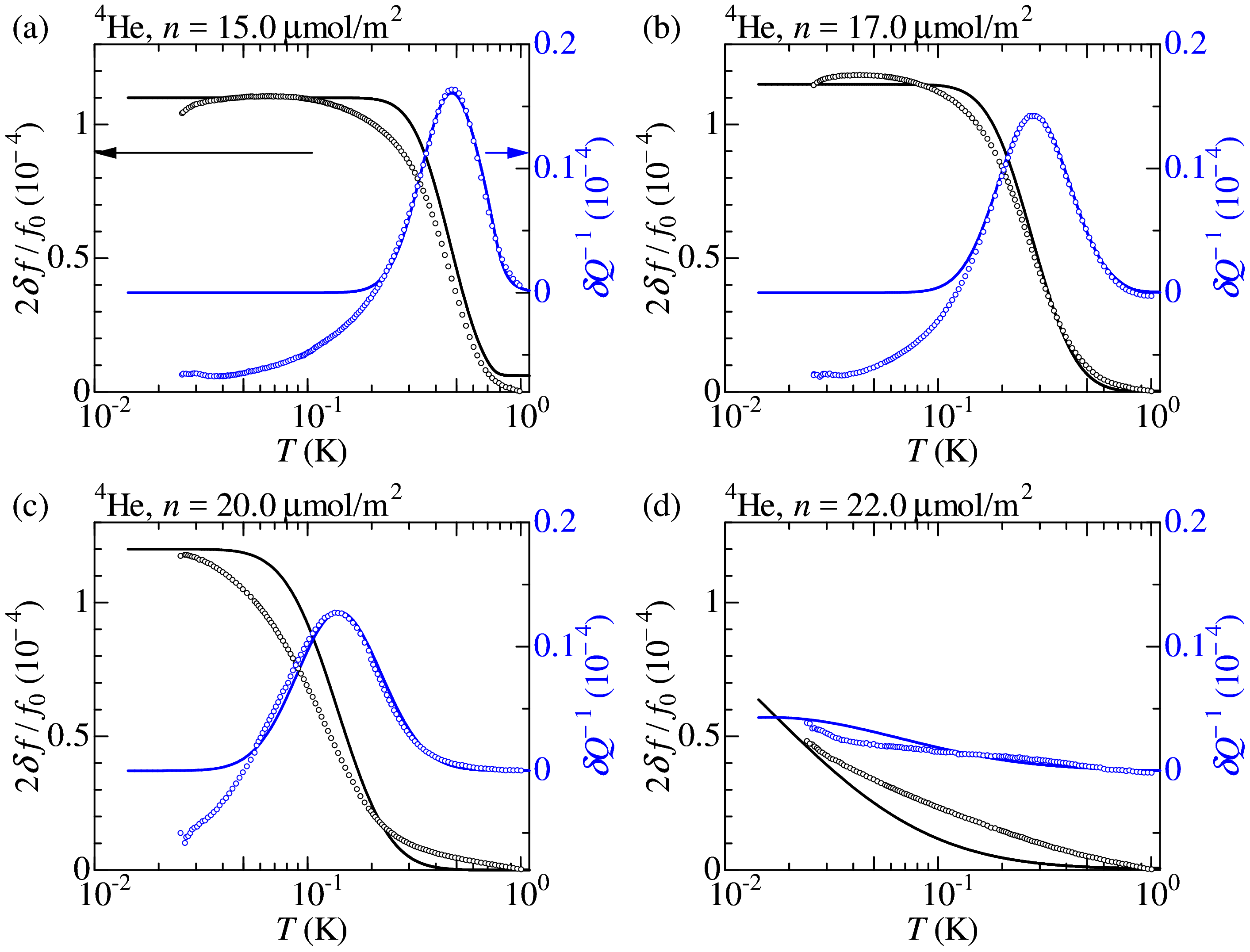}
  \caption{The normalized resonant frequency ${2\delta f}/{f_0}$ and excess dissipation $\delta Q^{-1}$ for $^4$He films at several coverages. Solid lines are the results of fitting to the complex response functions with a log-normal distributed energy gap (see text).
}
  \label{fig:data4He}
 \end{center}
\end{figure}

\begin{figure}[tbh]
 \begin{center}
  \centering
 \includegraphics[width=\columnwidth]{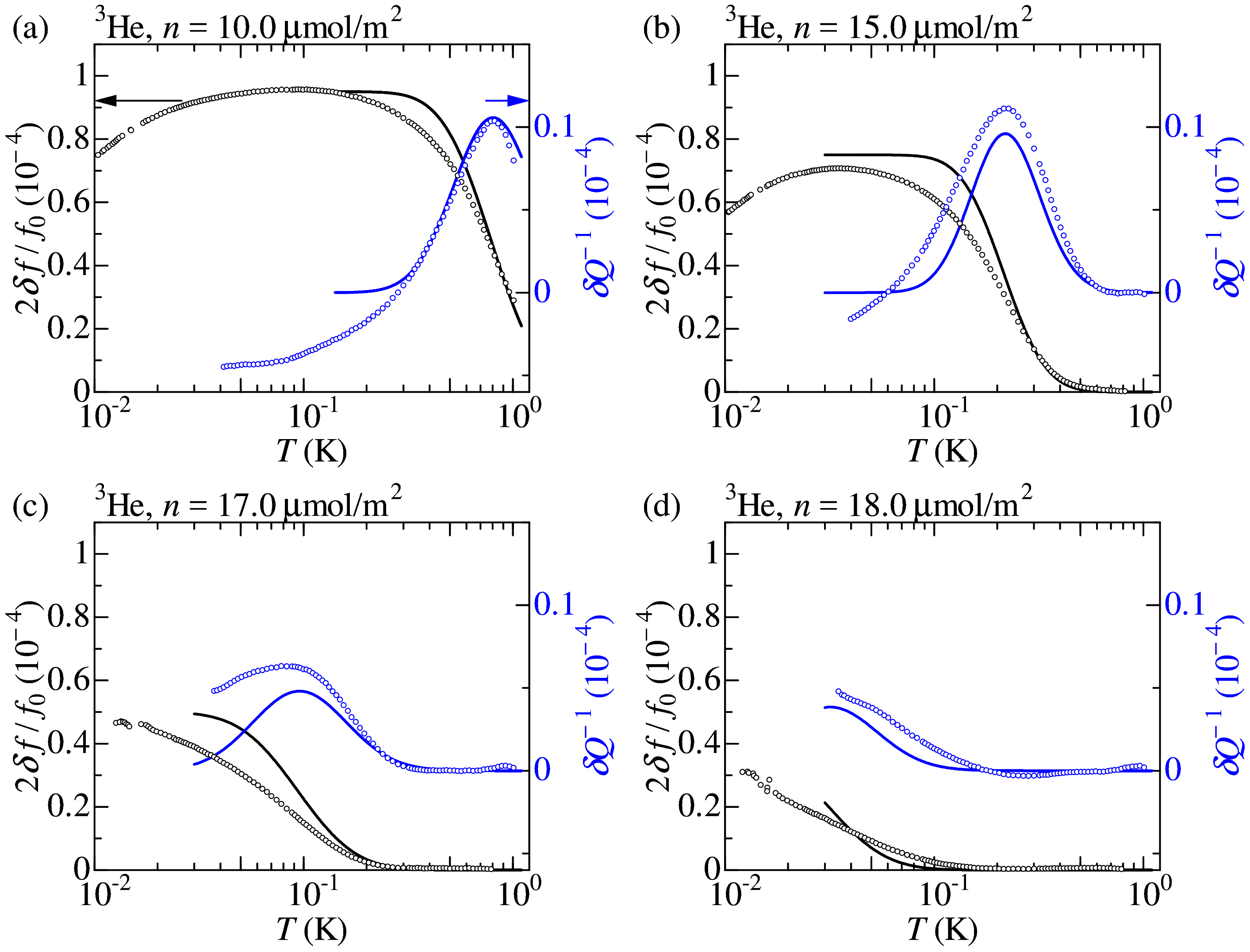}
  \caption{The normalized resonant frequency ${2\delta f}/{f_0}$ and excess dissipation $\delta Q^{-1}$ for $^3$He films at several coverages. Solid lines are the results of fitting to the complex response functions with a log-normal distributed energy gap (see text).}
  \label{fig:data3He}
 \end{center}
\end{figure}

\section{Interpretation of standard torsional oscillator experiments }
Our direct elasticity measurement was motivated by the observation of frequency shift and excess dissipation in a torsional oscillator for the study of superfluid properties of $^4$He films in porous Gelsil glass. %修正181014
Here we briefly discuss the results and interpretation in the previous TO studies. %was181014

Two TOs, which we refer to as TO1 and TO2, were employed as shown in Fig. \ref{fig:TO1+TO2}. %修正181014 
Each TO contained a disk sample of porous Gelsil glass inside the torsion bob. %修正181014 
In TO1, we glued all the faces of the glass sample to the wall by Stycast 1266 epoxy. The epoxy penetrated to the hole of the torsion rod was carefully removed by inserting a drill bit. 
On the other hand, in TO2, there was an open space between the porous glass and one side of the wall of the bob, at which the torsion rod is attached. % (actually the torsion rod and the bob wall is made of a single piece of BeCu).特に必要ないと思います181014

\begin{figure}[tb]
 \centering
 \includegraphics[width=\columnwidth]{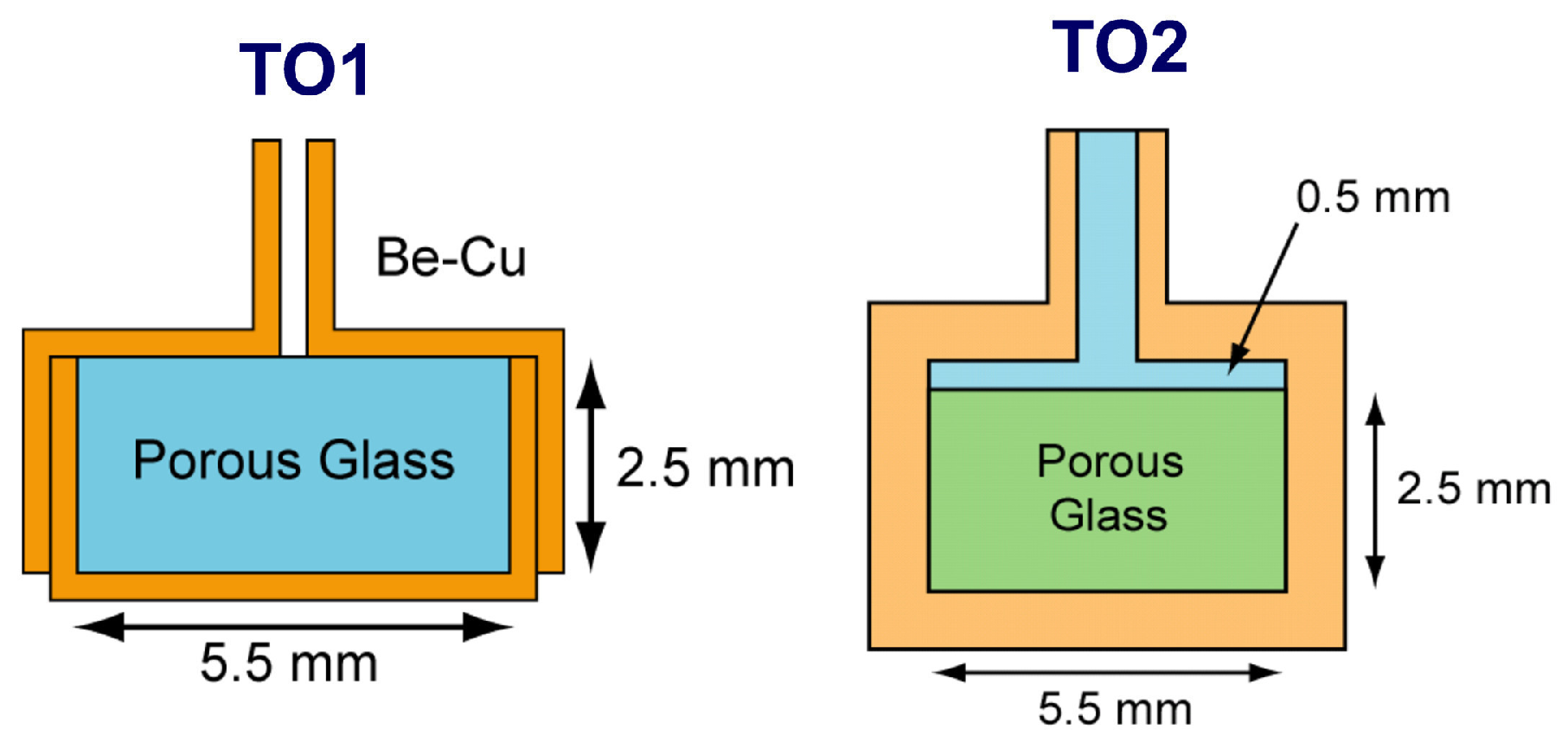}
 \caption{Schematic cross sectional views of TO1 and TO2.  %修正181014
The porous glass sample is glued to the BeCu enclosure with epoxy. }
 \label{fig:TO1+TO2}
\end{figure}

We performed measurements of $f$ and $Q^{-1}$ of TO1 and TO2 with adsorbed $^4$He at $ 6 < n < 35\ \mu$mol/m$^2$.
In TO1, we observed an increase in $f$ accompanied by a dissipation peak at $n < 22\ \mu$mol/m$^2$, as in the case of the present work. 
Figure \ref{fig:TO1-fQ} shows a result of fitting to the observation, which are converted to the normalized frequency shift $2\delta f/f_0$ and excess dissipation $\delta Q^{-1}$. 
We see that the overall $T$ dependencies of TO1 are identical to the results of the present TO, in which the Gelsil sample is located in torsion rod.
The fitting of the data to the complex response functions described in the main text works well.
We confirmed that in TO1 the dependence of the behaviors on $n$ and the obtained fitting parameters such as energy gap $\Delta$ are also identical to the present TO.
At $n > n_{\mathrm c}$, ordinary superfluid transitions were observed as an increase in $f$ below $T_{\mathrm c}$. %, which is shown in Fig. \ref{fig:phasediagram}. 181010Tc
%\textcolor{magenta}{
%The onset coverage of the elastic anomaly $T_\mathrm{onset}$,
The dissipation-peak temperature $T_\mathrm{p}$ and the superfluid transition temperature $T_\mathrm{c}$ are plotted in Fig. \ref{fig:phasediagramKogure}. 
The critical coverages $n_\mathrm{c}$ determined from the $n$ dependencies of $T_\mathrm{p}$ and $T_\mathrm{c}$ are identical within the accuracy of the data. %ここの記述are identical within the accuracy of the dataをもう少し改善できると良い181014 %本文の方を書き換えましたが，ここの改善は思いつきませんでした181015巻内
%結構です181016
%}

\begin{figure}[tb]
 \centering
 \includegraphics[width=80mm]{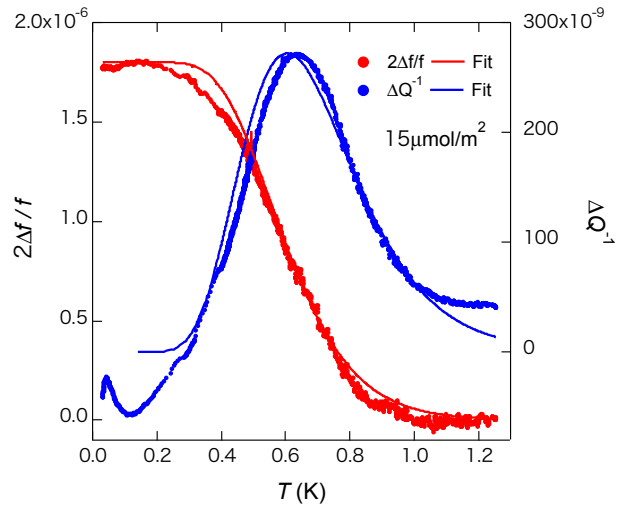}
 \caption{Anomalous response observed in TO1 for $^4$He coverage $n = 15.0\ \mu$mol/m$^2$. The data shown are after subtracting the background from the raw data and normalized to $2\delta f/f_0$ and $\delta Q^{-1}$ as in the main text. Note that the magnitude of effects is small by a factor of $10^{-2}$ compared with the present TO. Solid lines are the results of fitting similar to that described in the main text.}
 \label{fig:TO1-fQ}
\end{figure}

\begin{figure}[tb]
 \centering
 \includegraphics[width=60mm]{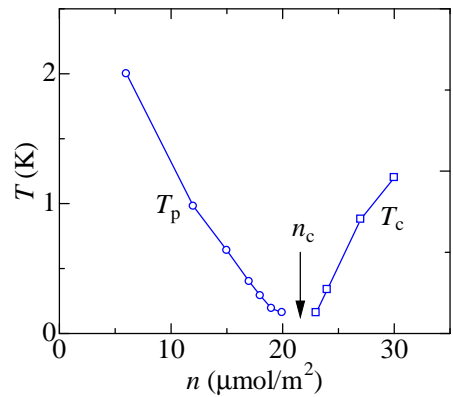}
 \caption{%\textcolor{magenta}{ [Referee B (2), NEW FIGURE]
 The dissipation-peak temperature $T_\mathrm{p}$ and the superfluid transition temperature $T_\mathrm{c}$ of $^4$He film detected with TO1.
 The arrow is at $n=21.6\ \mathrm{\mu mol/m^2}$.} %}
 \label{fig:phasediagramKogure}
\end{figure}

In TO2, however, such an elastic anomaly below $n_{\mathrm c}$ was not observed, while the superfluid transition was seen at $n > n_{\mathrm c}$ as in TO1. %修正181014
The superfluid transition temperature $T_\mathrm{c}$ in Fig. \ref{fig:phasediagram} is from TO2 \cite{Yamamoto2004}. %ref追加181015巻内
We have found that the absence of the elastic anomaly in TO2 is originated from the existence of open space between a face of porous glass disk and the wall of the TO cell near the torsion rod. 
We calculated the change in resonant frequency when the shear modulus of glass inside the TO bob increases, assuming the structures of TO1 and TO2 in FEM simulations.
The results are shown in Fig. \ref{fig:TO1TO2FEM}. 
When the shear modulus of glass inside TO1 increases 5 percent, $f$ increases about 60 mHz, while it increases only 2 mHz in TO2.

\begin{figure}[tb]
 \centering
 \includegraphics[width=70mm]{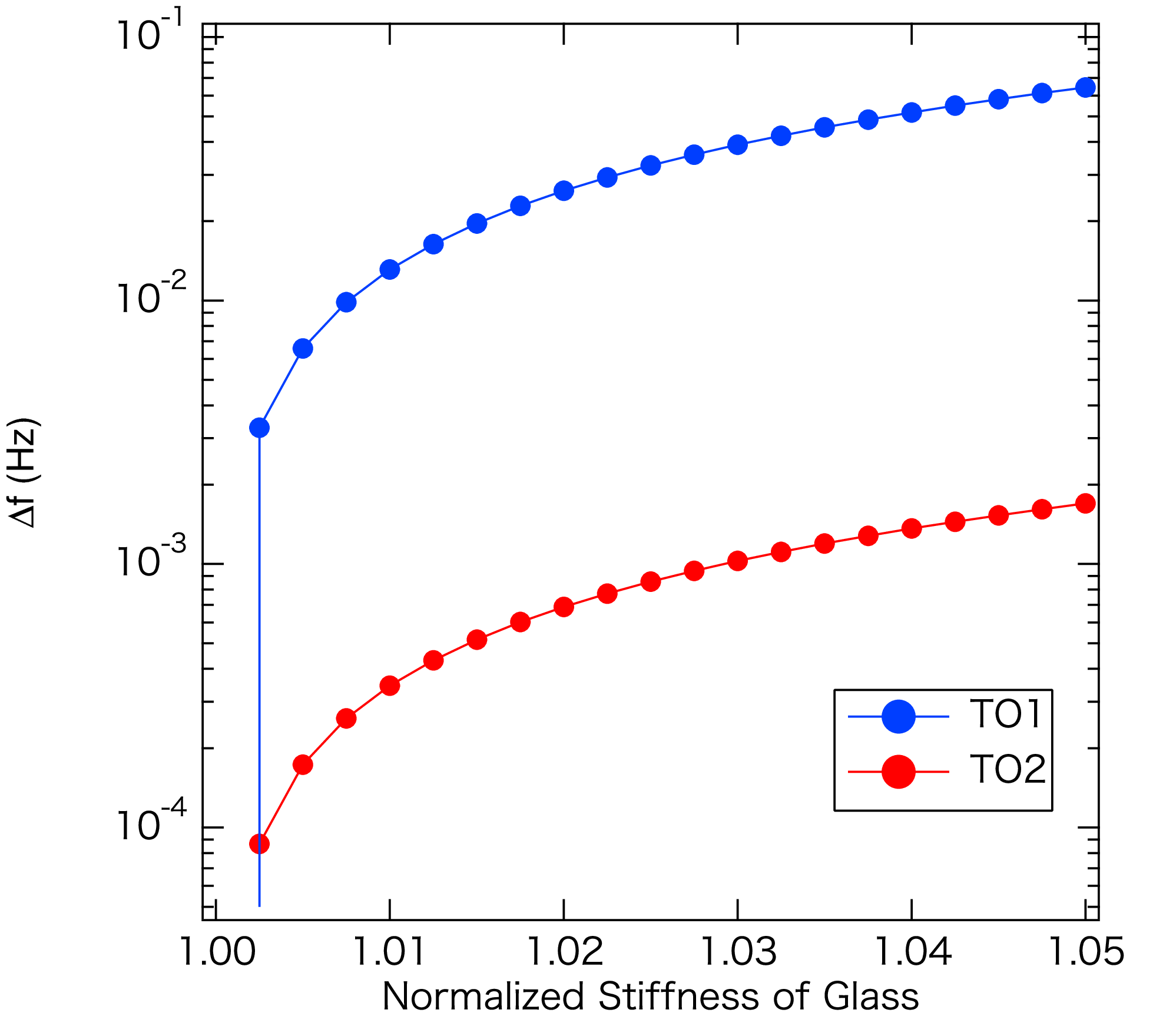}
 \caption{Calculated change in resonant frequency, $\delta f$, as a function of normalized stiffness of porous glass sample in TO1 and TO2: e.g. the value 1.01 corresponds to the one percent increase in shear modulus of glass by adsorbed helium. Difference of about 40 times between the cases of TO1 and TO2 is seen.}
 \label{fig:TO1TO2FEM}
\end{figure}

We interpret these results as follows: In a realistic TO made of metal for superfluid studies, %修正181014
the torsion bob is not rigid, and the resonant frequency of the fundamental torsion mode is determined not only by the shear modulus of the torsion rod but also by the shear modulus of the torsion bob, which consists of porous glass, BeCu enclosure and adsorbed helium in our experiments. 
This non-ideal nature of TO has been established by studies of apparent supersolidity of bulk solid $^4$He using TOs with many different designs\cite{Kim2004sci,Aoki2014,Kim2012}. %supersolid関連の引用必要181014 %このあたりを引用しましたが，どうでしょうか？181015巻内
%十分です181016
In particular, it has been realized as the Maris effect that the stiffness of the part of TO near the torsion rod has a large contribution to resonant frequency\cite{Maris2012}.
The presence of the elastic anomaly in TO1 and its absence in TO2 may be a manifestation of the Maris effect.
In TO2, the stiffening of porous glass sample by helium adsorption will hardly contribute to the total torsion constant by the existence of open space inside the bob.
We emphasize that this effect would be revealed only by FEM simulations, because it is difficult to calculate analytically the resonant frequency of a realistic TO with complicated structure and composites of different materials.

}%end of \appendix{}

%\nocite{*}


% The content of .bbl file is pasted below.
%

\providecommand{\noopsort}[1]{}\providecommand{\singleletter}[1]{#1}%

\end{document}